\documentclass[aps,prd,showpacs,superscriptaddress,nofootinbib,preprintnumbers,notitlepage]{revtex4-1} 
\usepackage{graphicx}
\usepackage{epsf}
\usepackage{bm}
\usepackage{amsmath}
\usepackage{amsfonts}
\usepackage{comment}
\usepackage{amssymb}
\usepackage{epstopdf}
\usepackage{natbib}
\usepackage{hyperref}
\usepackage{color}
\usepackage{verbatim}
\usepackage{multirow}
\usepackage{bm}
\usepackage{xcolor}
\usepackage{hyperref}
\usepackage[normalem]{ulem}
\definecolor{darkblue}{rgb}{0.0, 0.0, 0.55}
\definecolor{darkred}{rgb}{0.55, 0.0, 0.0}
\usepackage{hyperref}
\hypersetup{
    colorlinks=true,
    linkcolor=darkblue,
    citecolor=darkblue,
    urlcolor=darkblue}
    
\makeatletter\let\expandableinput\@@input\makeatother

\begin{document}

\title{Testing MOND-like modifications to gravity using growth-rate measurements and one-loop corrections to the matter power spectrum}

\author{Gabriel Hartmann}
\email{gabriel.hartmann@ufrgs.br}
\affiliation{Instituto de F\'{i}sica, Universidade Federal do Rio Grande do Sul, 91501-970 Porto Alegre RS, Brazil}

\author{Emanuelly Silva}
\email{emanuelly.santos@ufrgs.br}
\affiliation{Instituto de F\'{i}sica, Universidade Federal do Rio Grande do Sul, 91501-970 Porto Alegre RS, Brazil}

\author{Rafael C. Nunes}
\email{rafadcnunes@gmail.com}
\affiliation{Instituto de F\'{i}sica, Universidade Federal do Rio Grande do Sul, 91501-970 Porto Alegre RS, Brazil}
\affiliation{Divisão de Astrofísica, Instituto Nacional de Pesquisas Espaciais, Avenida dos Astronautas 1758, São José dos Campos, 12227-010, São Paulo, Brazil}

\begin{abstract}

We develop a perturbative framework for structure formation in a broad class of MOND-like theories characterized by a generalized nonlinear Poisson equation. We derive the modified evolution equations governing matter perturbations and obtain the corresponding linear growth equation, extending the analysis into the mildly nonlinear regime through one-loop corrections to the matter power spectrum. Beyond the theoretical framework, we perform a cosmological analysis based on two phenomenological scenarios: one parametrized by a quantity controlling the degree of nonlinearity in the generalized Poisson equation, and another describing the interplay between the MOND acceleration scale and the cosmological acceleration associated with the background expansion. We constrain these scenarios using recent measurements of the growth rate of structure, $f\sigma_8$, DESI-DR2 baryon acoustic oscillation data, and Type Ia supernova compilations. We find no statistically significant evidence for departures from the standard $\Lambda$CDM cosmology. The inferred constraints are fully consistent with the GR + $\Lambda$CDM scenario within the current observational uncertainties. At nonlinear scales, we note that MOND-like modifications can alter $P_{\rm NL}(k)$ and leave signatures that current and future high-precision large-scale-structure observations may probe. Our results establish a systematic connection between MOND-like gravitational dynamics and large-scale structure observations, providing a consistent framework to assess the phenomenological viability of MOND-inspired modifications of gravity in a cosmological context.

\end{abstract}

\maketitle

\section{Introduction}

Over the past decades, cosmological observations have seen extraordinary progress, enabling increasingly tight tests of the standard cosmological model and its underlying gravitational framework. Despite these paradigms' broad success, several statistically significant discrepancies have emerged as observational uncertainties have decreased. The most notable example is the $H_0$ tension, which concerns the discrepancy between local measurements of the cosmic expansion rate and the value inferred from cosmic microwave background (CMB) observations under $\Lambda$CDM \cite{DiValentino:2021izs}. 
The significance of this discrepancy currently exceeds $7\sigma$ \cite{H0DN:2025lyy}. In addition, several late-time probes have reported lower amplitudes of matter clustering than those preferred by early-universe observations, giving rise to the so-called $S_8$ tension \cite{Miyatake:2023njf, DES:2021wwk, Ivanov:2024xgb, Ivanov:2023qzb, Nunes:2021ipq, Benisty:2020kdt, DES:2026fyc} \footnote{It is important to emphasize that the interpretation of the $S_8$ tension depends on the adopted analysis methodology and dataset. In particular, the recent KiDS-Legacy cosmic shear analysis reduced the discrepancy with CMB measurements by approximately $2.3\sigma$ relative to the earlier KiDS-1000 result \cite{Wright:2025xka}. For a more detailed discussion of the origin of this shift and its implications, we refer the reader to \cite{Pantos:2026cxv}.} Beyond these two main ones, several other tensions and anomalies appear to persist within the current picture (see \cite{CosmoVerseNetwork:2025alb} for a comprehensive recent review).

% Although the origin of these discrepancies and anomalies remains uncertain, they have stimulated renewed interest in examining the fundamental assumptions underlying the standard cosmological framework. Since gravity governs both the expansion history of the Universe and the growth of cosmic structures, these tensions naturally raise the question of whether departures from the standard theory of gravity could leave observable imprints on cosmological data.

In addition to these cosmological tensions, the $\Lambda$CDM model also faces persistent challenges on galactic and subgalactic scales. Among the most widely discussed issues are the cusp--core problem, the missing satellites problem, the too-big-to-fail problem, and other small-scale discrepancies \cite{DelPopolo:2016emo, Kroupa:2012qj}. While astrophysical baryonic processes may alleviate some of these tensions, their ultimate origin remains an open question.

The observed mass discrepancy in galaxies and other gravitationally bound systems has also inspired the development of alternative theories of gravity, among which the Modified Newtonian Dynamics (MOND) paradigm stands as one of the most extensively studied proposals \cite{Milgrom:1983ca,Milgrom:1983zz,Bekenstein:2006bya,mond7}. Originally introduced as an alternative to particle dark matter, MOND successfully accounts for a variety of galactic-scale observations \cite{mond1,mond2,mond5,mond6,mond8,mond10,mond11,Zhang:2026oqg}. In particular, it naturally reproduces the approximately flat rotation curves of galaxies \cite{Lelli:2016zqa,mond3,mond9} and the baryonic Tully--Fisher relation \cite{mond4} through a modification of gravitational dynamics in the low-acceleration regime. For a comprehensive review of the MOND framework and its phenomenological implications, see Ref.\cite{Famaey:2011kh, Sanders:2002pf, Kroupa:2023ubo}.

However, extending MOND beyond galactic scales remains challenging. The modified dynamics alone is generally insufficient to account for the mass discrepancy observed in galaxy clusters and struggles to simultaneously reproduce several key cosmological and astrophysical observables, including the CMB, the growth of large-scale structure (LSS), and galaxy dynamics \cite{mond12,mond13, mond14,Rodrigues:2018duc,Marra:2020sts,Chan:2023zni,Chan:2022fcw,Rodrigues:2022oyd,Pradyumna:2021ufb}. 
These limitations have motivated the development of relativistic and covariant extensions, such as AQUAL \cite{Bekenstein:1984zz}, TeVeS \cite{Sanders_1997}, and several more recent formulations \cite{Skordis:2020eui,Chae:2024pxm,Hwang:2024rot,Kashfi:2022dyb,Domenech:2025qny, Bekenstein:2005nv, Skordis:2019fxt, Sebastiani:2016ras, Blanchet:2025wfr, Blanchet:2024mvy, Jacobson:2000xp}, which aim to provide a consistent description of gravitational phenomena across a wide range of scales. The variety of possibilities has grown. But assessing whether these theories are cosmologically viable still requires understanding how modified gravity affects structure formation.

In the standard framework, the continuity, Euler, and Poisson equations govern the evolution of matter perturbations and provide the basis for Standard Perturbation Theory (SPT) \cite{Bernardeau:2001qr}. In MOND-like scenarios, however, the Poisson equation is replaced by a generalized relation between the gravitational potential and the matter density field. This modification affects not only the linear growth of cosmic structures but also the nonlinear coupling between perturbation modes, leading to potentially observable signatures in LSS statistics.
The implications of such modifications for LSS observables within a consistent perturbative framework remain relatively unexplored. Establishing a systematic connection between MOND-inspired gravitational deformations and current cosmological observations is therefore an important step toward assessing their phenomenological viability.

\textit{The primary goal of this work is not to test gravity on cosmological scales in the absence of dark matter, nor to advocate MOND as a complete alternative to the standard cosmological model. Rather, we investigate whether MOND-inspired gravitational deformations, implemented on top of a $\Lambda$CDM background, provide a viable description of current cosmological observations and how their statistical performance compares with that of the standard scenario.}

Motivated by this perspective, we develop a perturbative framework for LSS formation in MOND-like theories. Adopting a phenomenological approach, we parametrize the modified gravitational dynamics through a generalized Poisson equation that recovers the Newtonian limit at high accelerations while allowing departures from standard gravity in the low-acceleration regime. We derive the corresponding modifications to the matter power spectrum and, within the effective field theory approach to large-scale structure, compute the associated one-loop corrections. Finally, we constrain the model parameters using recent cosmological observations, including Type Ia supernovae, the latest DESI DR2 baryon acoustic oscillation measurements, and a compilation of growth-rate data.

The paper is organized as follows. Section~\ref{sec:MOND-like} introduces the MOND-like framework and develops the perturbation theory for structure formation, emphasizing the linear regime and its cosmological implications. Section~\ref{sec:datasets} describes the observational datasets and the statistical methodology adopted in our analysis. Our main results are presented in Sec.~\ref{sec:constrains}. In Sec.~\ref{sec:eft} we extend our theoretical framework to the non-linear regime within the EFTofLSS. We conclude in Sec.~\ref{sec:end} with a summary of our findings and an outlook for future investigations.

%========================================================================================
\section{Linear structure formation in MOND-like gravity}
\label{sec:MOND-like}

We know that, within the SPT model, the evolution of the matter density contrast, $\delta(\mathbf{x}, \tau)$, and its peculiar velocity field, $\mathbf{v}(\mathbf{x}, \tau)$, in the Eulerian frame, is governed by the Euler, continuity, and Poisson equations \cite{Bernardeau:2001qr}:
\begin{equation}
\frac{\partial \mathbf{v}(\mathbf{x}, \tau)}{\partial \tau}
+ \mathcal{H}(\tau)\, \mathbf{v}(\mathbf{x}, \tau)
+ \big[\mathbf{v}(\mathbf{x}, \tau) \cdot \nabla\big] \mathbf{v}(\mathbf{x}, \tau)
= -\nabla \Phi(\mathbf{x}, \tau),
\label{eq:euler}
\end{equation}
\begin{equation}
\label{eq:cont}
\frac{\partial\delta(\mathbf{x,\tau})}{\partial \tau} + \nabla \cdot [(1+\delta(\mathbf{x},\tau))v(\mathbf{x},\tau)] = 0,
\end{equation}
\begin{equation}
\label{eq:Poissonn}
\nabla^2 \Phi(\mathbf{x}, \tau)
= \frac{3}{2}\, \mathcal{H}^2(\tau)\, \Omega_m(\tau)\, \delta(\mathbf{x}, \tau),
\end{equation}
where $\Phi$ is the gravitational potential, $\mathcal{H}$ is the conformal Hubble parameter, and $\Omega_m$ is the matter density parameter of the fiducial cosmological model. Physically, the first equation expresses momentum conservation, the second imposes mass conservation, and the third couples the scalar gravitational field to matter density perturbations.

In theories exhibiting MOND-like behavior, at field level, Eqs.~\eqref{eq:euler} and \eqref{eq:cont} remain unchanged, while the Poisson equation is replaced by a nonlinear relation between the gravitational potential and the matter density field. In a phenomenological description, in this work, let us consider the generalized Poisson equation (GPE) \cite{Evans1982ANP, Scherer:2025owa, Bekenstein:2004ne} 
\begin{equation}
\nabla \cdot \left[
F(s) \nabla\Phi(\mathbf{x},\tau)\right] = \frac{3}{2} \mathcal{H}^2(\tau) \Omega_m(\tau) \delta(\mathbf{x},\tau),
\label{GPE}
\end{equation}
with
\begin{equation}
F(s)=\left(\frac{s}{a_0}\right)^{p-2}.
\end{equation}
Here, $s=\left|\nabla\Phi(\mathbf{x},\tau)\right|$, $a_0$ denotes the characteristic MOND acceleration scale, and the parameter $p$ controls the degree of nonlinearity of the gravitational interaction. For $p>1$, Eq.~\eqref{GPE} defines a broad class of nonlinear gravitational theories \cite{Lindqvist:2019}. In particular, the standard Newtonian limit is recovered for $p=2$, for which Eq.~\eqref{GPE} reduces to the usual Poisson equation. On the other hand, for $p=3$, the gravitational potential outside the matter distribution becomes logarithmic, yielding the characteristic Milgromian potential in the low-acceleration regime.

This setup provides a convenient framework for testing departures from standard gravity, as the GPE encapsulates modifications to the gravitational interaction while preserving the fluid description encoded in the continuity and Euler equations. In this sense, it allows one to isolate the impact of modified gravity on structure formation without altering the standard matter conservation equations. Equipped with this phenomenological framework, we now investigate the implications of these modifications for the growth of cosmic structures.

To begin with, it is well known that the characteristic acceleration scale in MOND is empirically found to be of the order of
\begin{equation}
a_0 \sim 10^{-10} \, {\rm m\, s^{-2}}.
\end{equation}
From a cosmological perspective, however, the only natural acceleration scale associated with the background expansion is
\begin{equation}
a_{\rm cosmo}(z)=cH(z),
\end{equation}
where $H(z)$ is the Hubble parameter and $c$ is the speed of light. Remarkably, at the present epoch this quantity is of the same order of magnitude as the MOND acceleration scale,
\begin{equation}
a_0 \sim cH_0 \sim 10^{-10} \, {\rm m\,s^{-2}}.
\end{equation}
\textit{This intriguing numerical coincidence has long been regarded as a possible indication of a deeper connection between the MOND acceleration scale and cosmology}. Indeed, it suggests that the characteristic acceleration governing the transition between the Newtonian and MOND regimes may not be a fundamental constant, but rather an emergent scale related to the cosmological background evolution. Motivated by this possibility, we will also explore a phenomenological extension in which the MOND acceleration scale is allowed to depend on the cosmological expansion rate.

We now adopt a consistent approach to linearize Eq.~\eqref{GPE}. Since the generalized Poisson equation is intrinsically nonlinear, a perturbative treatment requires specifying a background gravitational environment around which the equations can be expanded. Following the external-field approximation, commonly employed in MOND-like theories, we assume that the total gravitational potential can be decomposed into a background contribution plus a perturbation,
\begin{equation}
\Phi(\mathbf{x},\tau)=\Phi_{*}(\mathbf{x},\tau)+\varphi(\mathbf{x},\tau),
\end{equation}
where the background field is assumed to vary slowly over the scales of interest and can therefore be approximated as spatially uniform, such that
\begin{equation}
\nabla\Phi_{*}(\mathbf{x},\tau)\equiv\mathbf{g}_{0},
\end{equation}
with $\mathbf{g}_{0}$ being a constant vector. Consequently, the total gravitational field can be written as
\begin{equation}
\nabla\Phi(\mathbf{x},\tau)=\mathbf{g}_{0}+\nabla\varphi(\mathbf{x},\tau),
%\qquad
%\nabla^{2}\Phi(\mathbf{x},\tau)=\nabla^{2}\varphi(\mathbf{x},\tau),
\end{equation}
%where we have used the fact that $\nabla\cdot\mathbf{g}*{0}=0$. 
We further assume that the perturbation is small compared to the background field,
\begin{equation}
|\nabla\varphi(\mathbf{x},\tau)|\ll|\mathbf{g}_{0}|.
\end{equation}
Defining $g_{0}\equiv |\mathbf{g}_{0}|$ and recalling that $s=|\nabla\Phi(\mathbf{x},\tau)|$, we obtain

\begin{equation}
s=|\mathbf{g}_{0}+\nabla\varphi(\mathbf{x},\tau)|
%=\sqrt{\left(\mathbf{g}_{0}+\nabla\varphi(\mathbf{x},\tau)\right)\cdot\left(\mathbf{g}_{0}+\nabla\varphi(\mathbf{x},\tau)\right)}
=g_{0}\sqrt{1+2\frac{\mathbf{g}_{0}\cdot\nabla\varphi(\mathbf{x},\tau)}{g_{0}^{2}}+\frac{|\nabla\varphi(\mathbf{x},\tau)|^{2}}{g_{0}^{2}}}.
\end{equation}

Since $|\nabla\varphi(\mathbf{x},\tau)|\ll g_{0}$, the last term is second order in the perturbation and can be neglected. 
%Expanding the square root to first order according to $\sqrt{1+x}\simeq1+x/2$ for $|x|\ll1$, we finally obtain
Expanding the square root to first order, we obtain
\begin{equation}
s\simeq g_{0}+\frac{\mathbf{g}_{0}\cdot\nabla\varphi(\mathbf{x},\tau)}{g_{0}}.
\label{eq:slinear}
\end{equation}
Equation~\eqref{eq:slinear} provides the linearized expression for the magnitude of the gravitational field and constitutes the starting point for deriving the perturbative form of the generalized Poisson equation.

Now that we have obtained an approximate expression for the variable $s$, we can Taylor expand the function
\begin{equation}
F(s)=a_0^{-(p-2)}s^{p-2}
\end{equation}
around the background value $g_0$, yielding
\begin{equation}
F(s)
\simeq
F(g_0)
+
F'(g_0)
\frac{\mathbf g_0\cdot\nabla\varphi(\mathbf{x},\tau)}
{g_0}.
\end{equation}

Defining
\begin{equation}
F_0\equiv F(g_0)=a_0^{-(p-2)}g_0^{p-2},
\end{equation}
and
\begin{equation}
F'_0\equiv F'(g_0)=(p-2)a_0^{-(p-2)}g_0^{p-3},
\end{equation}
we obtain
\begin{equation}
F(s)
\simeq
F_0
+
F'_0
\frac{\mathbf g_0\cdot\nabla\varphi(\mathbf{x},\tau)}
{g_0}.
\end{equation}

Substituting this expansion into the generalized Poisson equation eq. (\ref{GPE}),
%\begin{equation}
%\nabla \cdot \left[
%F(s)\nabla\Phi(\mathbf{x},\tau)
%\right]
%=
%\frac{3}{2}
%\mathcal{H}^2(\tau)
%\Omega_m(\tau)
%\delta(\mathbf{x},\tau),
%\end{equation}
%and using
%\begin{equation}
%\nabla\Phi=\mathbf g_0+\nabla\varphi,
%\end{equation}
while neglecting terms of order $\mathcal{O}(|\nabla\varphi|^2)$, we obtain 
\begin{equation}
\nabla \cdot \left[F(s)\nabla\Phi\right]
=
F_0 \nabla^2 \varphi(\mathbf{x},\tau)
+
F'_0 \frac{(\mathbf g_0\cdot\nabla)^2}{g_0}\varphi(\mathbf{x},\tau).
\end{equation}

Therefore, the linearized generalized Poisson equation becomes
\begin{equation}
F_0 \nabla^2 \varphi(\mathbf{x},\tau)
+
F'_0 \frac{(\mathbf g_0\cdot\nabla)^2}{g_0}\varphi(\mathbf{x},\tau)
=
\frac{3}{2}
\mathcal{H}^2(\tau)
\Omega_m(\tau)
\delta(\mathbf{x},\tau),
\label{eq:linearGPE}
\end{equation}
which is valid in the regime $|\nabla\varphi(\mathbf{x},\tau)| \ll g_0$.

In this work, we adopt the Fourier convention
\begin{equation}
f(\mathbf{x})
=
\int \frac{d^3k}{(2\pi)^3}
e^{i\mathbf{k}\cdot\mathbf{x}}
f(\mathbf{k}),
\qquad
f(\mathbf{k})
=
\int d^3x \,
e^{-i\mathbf{k}\cdot\mathbf{x}}
f(\mathbf{x}).
\label{fourier}
\end{equation}
%such that
%\begin{equation}
%\nabla^2 \rightarrow -k^2,
%\qquad
%(\mathbf g_0\cdot\nabla)^2 \rightarrow -(\mathbf g_0\cdot\mathbf k)^2.
%\end{equation}

In Fourier space, Eq.~\eqref{eq:linearGPE} becomes
\begin{equation}
-F_0 k^2 \varphi(\mathbf{k},\tau)
-
F'_0 \frac{(\mathbf g_0\cdot\mathbf k)^2}{g_0}
\varphi(\mathbf{k},\tau)
=
\frac{3}{2}
\mathcal{H}^2(\tau)
\Omega_m(\tau)
\delta(\mathbf{k},\tau).
\end{equation}

Finally, solving for the potential yields
\begin{equation}
\varphi(\mathbf{k},\tau)
=
-
\frac{
\frac{3}{2}
\mathcal{H}^2(\tau)
\Omega_m(\tau)
}{
F_0 k^2
+
\frac{F'_0}{g_0}(\mathbf g_0\cdot\mathbf k)^2
}
\,
\delta(\mathbf{k},\tau).
\label{eq:phiFourier}
\end{equation}

We now focus on the linear regime of structure formation. In this limit, density perturbations are assumed to be small, $|\delta|\ll 1$, allowing us to neglect nonlinear terms in the equations of motion. Under this approximation, the fundamental equations of SPT in our framework can be written as
\begin{equation}
\frac{\partial \mathbf{v}(\mathbf{x},\tau)}{\partial \tau}
+
\mathcal{H}(\tau)\mathbf{v}(\mathbf{x},\tau)
=
-\nabla\Phi(\mathbf{x},\tau),
\end{equation}
\begin{equation}
\frac{\partial\delta(\mathbf{x},\tau)}{\partial \tau}
+
\nabla\cdot\mathbf{v}(\mathbf{x},\tau)
=0,
\end{equation}
\begin{equation}
\nabla^{2}\varphi(\mathbf{x},\tau)
=
\frac{3}{2}\mathcal{H}^{2}(\tau)\Omega_{m}(\tau)\delta(\mathbf{x},\tau).
\label{eq:Poisson}
\end{equation}
These correspond, respectively, to the Euler equation, the continuity equation, and the Poisson equation, which together govern the evolution of the matter density contrast and velocity field in an expanding background.
We define the velocity divergence as $\theta(\mathbf{x},\tau)\equiv\nabla\cdot\mathbf{v}(\mathbf{x},\tau)$, which allows us to rewrite the system in terms of scalar quantities only. Taking the divergence of the Euler equation, in Fourier space, we obtain
%\begin{equation}
%\frac{\partial\theta(\mathbf{x},\tau)}{\partial\tau}
%+
%\mathcal{H}(\tau)\theta(\mathbf{x},\tau)
%=
%-\nabla^{2}\Phi(\mathbf{x},\tau).
%\end{equation}

%It is convenient to move to Fourier space, where spatial derivatives become algebraic operations. In this work, we adopt the Fourier convention
%\begin{equation}
%f(\mathbf{x})
%=
%\int \frac{d^{3}k}{(2\pi)^{3}}
%e^{i\mathbf{k}\cdot\mathbf{x}}f(\mathbf{k}),
%\qquad
%f(\mathbf{k})
%=
%\int d^{3}x\,e^{-i\mathbf{k}\cdot\mathbf{x}}f(\mathbf{x}),
%\end{equation}
%such that the previous equation becomes
\begin{equation}
\frac{\partial\theta(\mathbf{k},\tau)}{\partial\tau}
+
\mathcal{H}(\tau)\theta(\mathbf{k},\tau)
=
-k^{2}\Phi(\mathbf{k},\tau).
\end{equation}

We now use the continuity equation to eliminate $\theta(\mathbf{k},\tau)$ in favor of the density contrast. Differentiating the continuity equation with respect to conformal time and combining it with the Euler equation yields
\begin{equation}
\frac{\partial^{2}\delta(\mathbf{k},\tau)}{\partial\tau^{2}}
+
\mathcal{H}(\tau)\frac{\partial\delta(\mathbf{k},\tau)}{\partial\tau}
=
-k^{2}\Phi(\mathbf{k},\tau).
\end{equation}

Substituting $\varphi(\mathbf{k},\tau)$ from the modified Poisson relation, we obtain a closed second-order differential equation describing the linear evolution of each Fourier mode $k$. This equation determines the growth of matter perturbations in the linear regime:
\begin{equation}
\frac{\partial^{2}\delta(\mathbf{k},\tau)}{\partial\tau^{2}}
+
\mathcal{H}(\tau)\frac{\partial\delta(\mathbf{k},\tau)}{\partial\tau}
-
\frac{3}{2}\mathcal{H}^{2}(\tau)\Omega_{m}(\tau)
\frac{k^{2}}
{F_{0}k^{2}+\frac{F'_{0}}{g_{0}}(\mathbf g_{0}\cdot\mathbf k)^{2}}
%_{\mu_{\rm eff}(\mathbf{k},\tau)}
\delta(\mathbf{k},\tau)
=0.
\end{equation}

We can further simplify the effective coupling by introducing $\mu\equiv\cos\theta=(\mathbf g_{0}\cdot\mathbf k)/(g_{0}k)$, where $\theta$ is the angle between $\mathbf g_{0}$ and $\mathbf k$. Using $F_{0}=a_{0}^{-(p-2)}g_{0}^{p-2}$ and $F'_{0}=(p-2)a_{0}^{-(p-2)}g_{0}^{p-3}$, we obtain
\begin{equation}
\label{growth_geral}
\frac{\partial^{2}\delta(\mathbf{k},\tau)}{\partial\tau^{2}}
+
\mathcal{H}(\tau)\frac{\partial\delta(\mathbf{k},\tau)}{\partial\tau}
-
\frac{3}{2}\mathcal{H}^{2}(\tau)\Omega_{m}(\tau)
\left(\frac{a_{0}}{g_{0}}\right)^{p-2}
\frac{1}{1+(p-2)\mu^{2}}
\delta(\mathbf{k},\tau)
=0.
\end{equation}

We define the effective coupling as
\begin{equation}
\label{coupling_mond}
\mu_{\rm eff}(\mathbf{k},\tau)
=
\left(\frac{a_{0}}{g_{0}}\right)^{p-2}
\frac{1}{1+(p-2)\mu^{2}}.
\end{equation}

This is the linear second-order equation for the dark-matter density contrast. From this equation it is clear that when $\mu_{\rm eff} > 1$, the effective gravitational strength is enhanced relative to the standard case, leading to a growth of perturbations, whereas for $\mu_{\rm eff} < 1$ the growth is suppressed with respect to $\Lambda$CDM.

In the linear regime, $\delta(\mathbf{x},\tau)$ appearing in Eq.~(\ref{growth_geral}) can be factorized in terms of the linear growth function $D(\tau)$, which characterizes the time evolution of density perturbations driven by gravitational instability. It is defined through
\begin{equation}
\delta(\mathbf{x},\tau)=D(\tau)\,\delta(\mathbf{x},\tau_0),
\qquad
D(\tau)=\frac{\delta(\mathbf{x},\tau)}{\delta(\mathbf{x},\tau_0)},
\label{eq:gf}
\end{equation}
where $\delta(\mathbf{x},\tau_0)$ denotes the density contrast evaluated at a reference time, typically chosen to correspond to the present epoch. The growth function is conventionally normalized such that $D(z=0)=1$.

With this definition, Eq.~(\ref{growth_geral}) becomes
\begin{equation}
\frac{d^2D(\tau)}{d\tau^2}
+
\mathcal{H}(\tau)\frac{dD(\tau)}{d\tau}
-
\frac{3}{2}\mathcal{H}^2(\tau)\Omega_m(\tau)
\mu_{\rm eff}(k,\mu)
D(\tau)
=0,
\label{eq:growth}
\end{equation}
where $\mu_{\rm eff}(k,\mu)$ is given by Eq.(\ref{coupling_mond}).

For $p=2$, one has $\mu_{\rm eff}=1$, and the standard $\Lambda$CDM growth equation is recovered. The effective modification $\mu_{\rm eff}$ depends on the relative orientation between the Fourier mode $\mathbf{k}$ and the external field $\mathbf{g}_0$, introducing an anisotropic correction to the growth of structure. These results demonstrate how MOND-like dynamics affect structure formation even in the linear regime, leading to scale- and direction-dependent modifications of the growth rate.
\\

\begin{comment}

The characteristic acceleration scale $a_0$ controls the transition between the standard (Newtonian) and modified regimes, while the dimensionless ratio $\frac{a_0}{g_0}$ quantifies how deeply the system probes the nonlinear dynamics. Since this ratio fully determines the anisotropic corrections to the growth of cosmological perturbations, it is natural to associate the external field with the Hubble expansion rate, i.e. $g_0 = c\,H(z)$. Under this assumption, the ratio $\frac{a_0}{g_0}$ can be written as a redshift-dependent function,
\begin{equation}
    \frac{a_0}{g_0}(z) = \frac{H_0}{H(z)}.
\end{equation}

In general lines, one can propose that 

\begin{equation}
    \frac{a_0}{g_0}(z) = \left(\frac{H_0}{H(z)}\right)^m,
\end{equation}
where the parameter $m$ control ............

Another import feature is that we replace $\mu_{eff}(k,\tau)$ by it is angular average $\langle\mu_{\rm  eff}\rangle(\tau)$, such that:
\begin{equation}
    \langle\mu_{\rm eff}\rangle(\tau) = \frac{1}{2}\int_{-1}^{1} \mu_{\rm  eff}(k,\tau) \space d\mu
\end{equation}

\end{comment}

The characteristic acceleration scale $a_0$ controls the transition between the standard (Newtonian) and modified regimes, while the dimensionless ratio $a_0/g_0$ quantifies how deeply the system probes the nonlinear dynamics. Since this ratio fully determines the amplitude of the anisotropic corrections to the growth of cosmological perturbations, it is natural to associate the external field with the Hubble expansion rate, i.e. $g_0 = c\,H(z)$. Under this assumption, the ratio $a_0/g_0$ can be written as a redshift-dependent function,
\begin{equation}
\frac{a_0}{g_0}(z)=\frac{a_0}{cH(z)}.
\end{equation}

In general, one may consider a more flexible phenomenological parametrization,
\begin{equation}
\frac{a_0}{g_0}(z)=\left(\frac{H_0}{H(z)}\right)^m,
\end{equation}
where the parameter $m$ controls the deviation from the simple linear scaling and encodes possible departures from a direct identification between the MOND acceleration scale and the cosmological expansion rate.

It is important to emphasize that the identification $g_0 = cH(z)$ should be understood as a phenomenological ansatz rather than a fundamental prediction of the theory. In the present framework, $g_0$ represents an external-field scale controlling the transition between the linear and nonlinear regimes of the generalized Poisson equation. The assumption that this scale tracks the cosmological expansion rate is motivated by the well-known numerical coincidence between the MOND acceleration scale and the present-day Hubble scale, $a_0 \sim cH_0$, which has been extensively discussed in the literature. From this perspective, the proposed scaling provides a minimal and testable way to incorporate a possible cosmological evolution of the MOND transition scale, allowing one to assess whether large-scale structure data favor a direct connection between local gravitational dynamics and the background expansion history. Importantly, this choice does not follow from the underlying field equations and should be regarded as an effective parametrization to be constrained observationally.

Another important feature is that we replace $\mu_{\rm eff}(k,\tau)$ by its angular average $\langle\mu_{\rm eff}\rangle(\tau)$, defined as
\begin{equation}
\langle\mu_{\rm eff}\rangle(\tau)
=
\frac{1}{2}\int_{-1}^{1}\mu_{\rm eff}(k,\mu,\tau)\,d\mu,
\end{equation}
which removes the explicit anisotropy associated with the preferred direction defined by $\mathbf{g}_0$ and allows for a statistically isotropic description of structure formation.

Having established the main equations, we now turn our attention to the impact on structure formation, focusing in particular on the linear regime. In order to quantify the effects of MOND-like modifications on the evolution of matter perturbations, it is essential to compare theoretical predictions with cosmological observables. A key example is redshift-space distortions (RSD), which arise from velocity-induced effects when mapping galaxy positions from real space to redshift space due to peculiar motions along the line of sight. These distortions generate anisotropies in the observed clustering pattern and are directly sensitive to the growth of cosmic structures. As a result, RSD measurements constrain the combination $f\sigma_8(z)$, or equivalently $f(a)\sigma_8(a)$, where $\sigma_8(a)$ denotes the variance of the matter density field smoothed on a scale of $R=8\,h^{-1}\mathrm{Mpc}$, while $f(a)$ is the logarithmic growth rate defined as
\begin{equation}
f(a)=\frac{d\ln D(a)}{d\ln a}.
\end{equation}

In this context, the modified growth equation introduces a scale- and direction-dependent effective gravitational coupling through the function $\mu_{\rm eff}(k,\mu,\tau)$. As a consequence, both the growth rate $f(a)$ and the amplitude of matter fluctuations $\sigma_8(a)$ acquire implicit dependencies on the wavenumber and on the relative orientation between $\mathbf{k}$ and the external field direction $\mathbf{g}_0$. In particular, the anisotropic structure of $\mu_{\rm eff}(k,\mu,\tau)$ implies that the linear growth of perturbations is no longer fully separable into time- and scale-dependent components in a strictly isotropic sense.

For redshift-space distortions, this has direct implications, since the observed quantity $f\sigma_8(z)$ is effectively sensitive to the angularly averaged growth of structure. In practice, the presence of the preferred direction $\mathbf{g}_0$ introduces an anisotropic modulation of the growth factor, which must be averaged over orientations to obtain a quantity that can be compared with isotropic RSD measurements. In this work, we therefore interpret the theoretical prediction for $f\sigma_8$ as arising from an effective growth rate constructed from the angularly averaged coupling $\langle \mu_{\rm eff}(k,\tau)\rangle$.

This averaging procedure allows us to map the inherently anisotropic linear dynamics of MOND-like theories onto the standard isotropic observables probed by galaxy surveys. As a result, deviations from $\Lambda$CDM in RSD data can be directly traced back to the parameters $(p, a_0/g_0)$ controlling both the strength and the anisotropy of the modified gravitational interaction, providing a clear observational handle to constrain MOND-like departures from standard structure formation.

\begin{figure}[htpb!]
    \centering
    \includegraphics[scale=0.492]{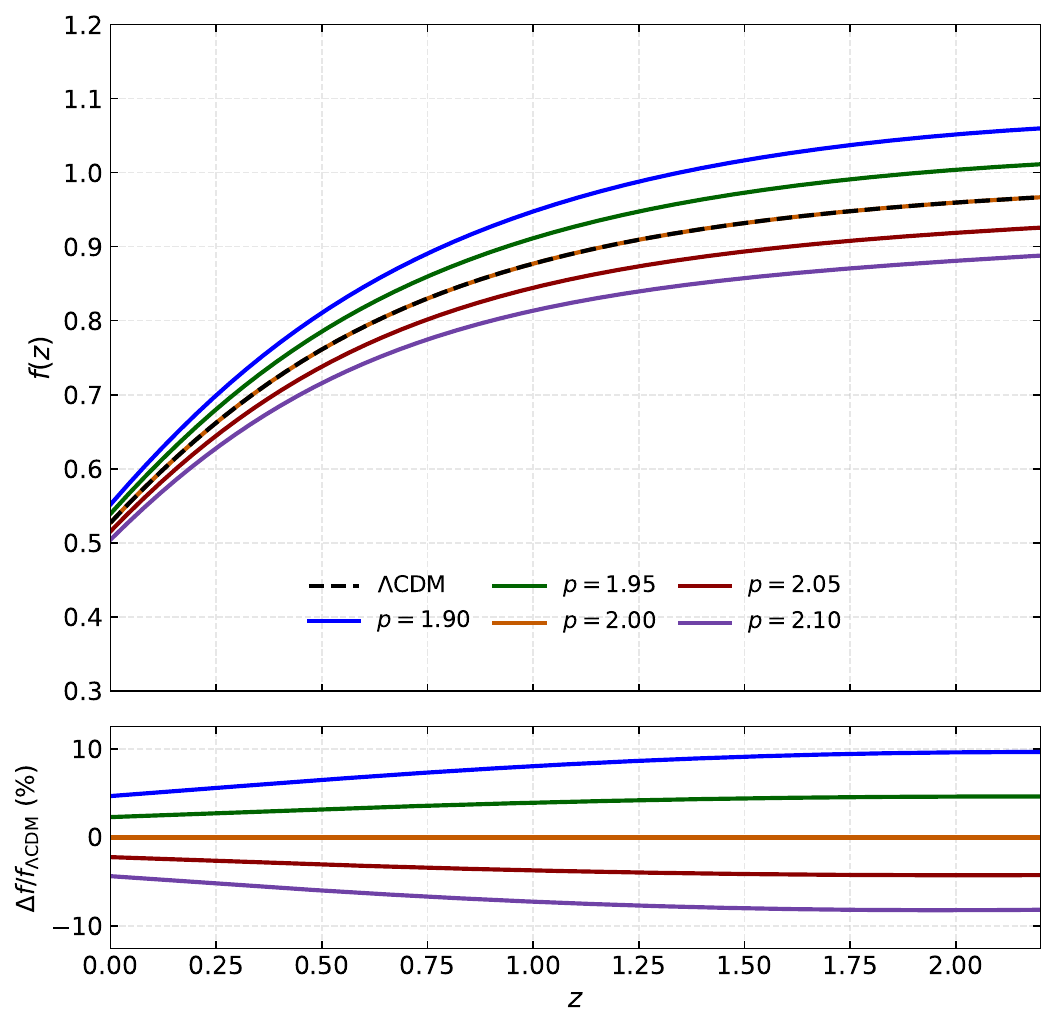}
    \includegraphics[scale=0.499]{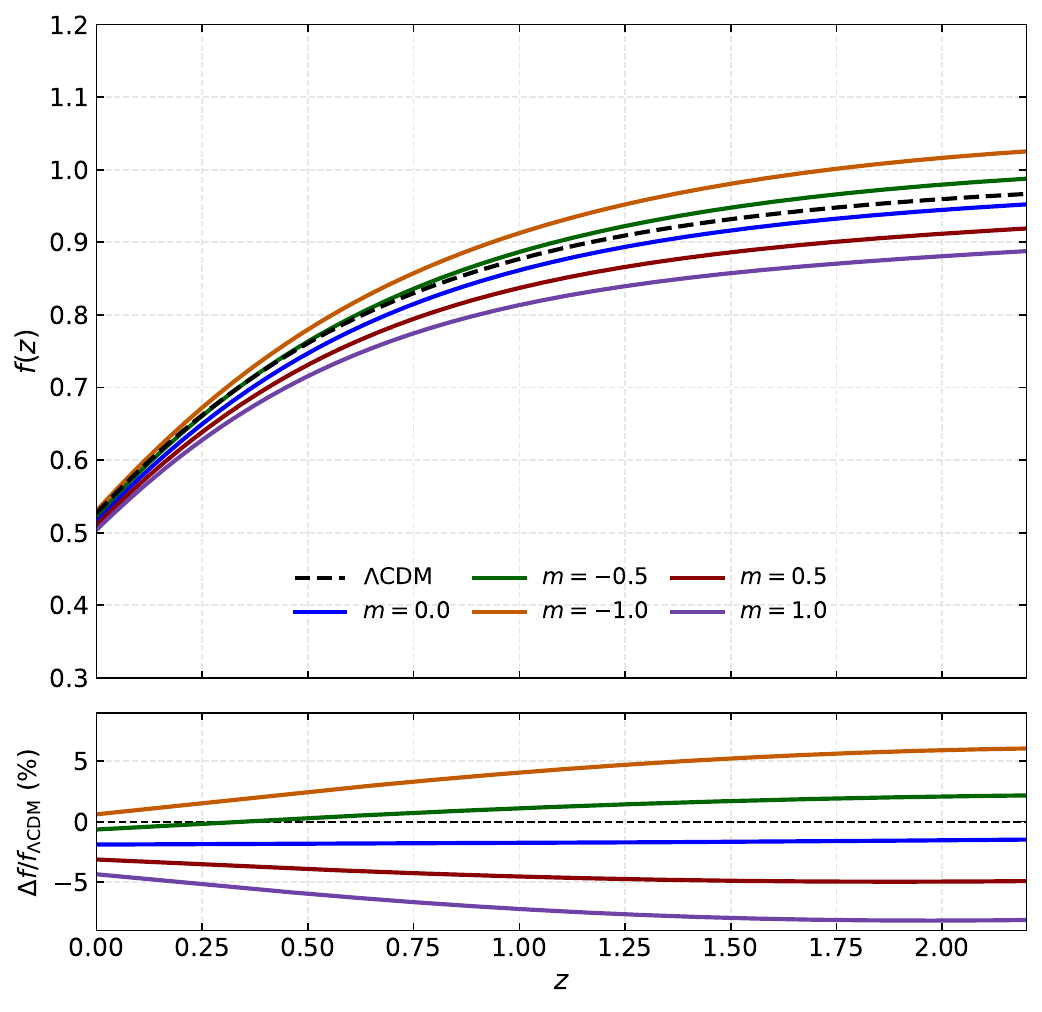}
    \caption{\textit{Left panel:} Theoretical predictions for $f(z)$ obtained for different values of $p$ (see legend). The black dashed curve corresponds to the best-fit $\Lambda$CDM model. The lower subpanel shows the relative difference of each model with respect to the $\Lambda$CDM prediction. \textit{Right panel:} Same as the left panel, but with $p=2.10$ fixed while varying only $m$. Note that the case $m=0$ does not reduce to $\Lambda$CDM, since additional contributions arising from the choice $p\neq2$ are still present.}
    \label{fig:f_z}
\end{figure}

Figure \ref{fig:f_z} shows the theoretical prediction of MOND-like models for the observable quantity $f(z)$. In both the left panel (where $m=1$ is fixed) and the right panel (where $p=2.10$ is fixed), we observe that the difference between the MOND-like dynamics and $\Lambda$CDM becomes more pronounced towards higher redshifts. This occurs because the effective gravitational coupling, $\mu_{\rm eff}$, which is directly related to the parameters $p$ and $m$, depends on the expansion history through the ratio $H_0/H(z)$. Since $H(z)$ is larger at earlier times, the modified gravitational dynamics become more pronounced. Regarding the trends observed, we find in the left panel that larger values of $p$ supress the growth rate relative to $\Lambda$CDM, whereas smaller values increase its amplitude. With $p$ held fixed, varying $m$ produces a smaller deviation form $\Lambda$CDM, although the same qualitative trend remains.

These results show that the modification in Equation~\eqref{GPE} can substantially affect the growth of cosmic structures. This is important because it means that changes in the model parameters can alter the predicted amplitude of matter clustering and the inferred value of $S_8$. This may reduce the discrepancy between early- and late-Universe measurements, potentially easing the $S_8$ tension.

\section{Datasets and Methodology}
\label{sec:datasets}

In the following, we describe the numerical methodology employed in our data analysis, together with the observational datasets considered in this work.

\subsection{\texttt{SERTAO} code\footnote{https://github.com/rafaelcnunes/SERTAO}}

To rigorously test our proposed theoretical framework, we implemented it within the \texttt{SERTAO} analysis pipeline, a fully self-consistent cosmological inference framework developed to perform background and late-time observational constraints in a flexible and modular manner. \texttt{SERTAO} compute the background expansion history and all relevant late-time observables in a fully self-consistent manner. The cosmological evolution is solved numerically through a dedicated background module that directly integrates the modified Friedmann equations and computes derived quantities such as $D_A(z)$, $D_L(z)$, and $H(z)$ without relying on external Boltzmann solvers. In addition to the background evolution, the \texttt{SERTAO} framework includes an independent module for the computation of linear structure formation. The growth of matter perturbations is obtained by numerically integrating the linear perturbation equations directly from the modified cosmological background, allowing us to compute the evolution of the growth factor $D(z)$, the growth rate $f(z)$, and derived observables such as $f\sigma_8$. This implementation ensures full consistency between the background dynamics and the perturbative sector within the same theoretical assumptions.

Importantly, this perturbation module is implemented independently of external Boltzmann solvers, providing a transparent and flexible environment to test generalized cosmological scenarios. The numerical integration is stable across the relevant redshift range and allows direct comparison with large-scale structure observables without relying on pre-tabulated transfer functions.

Furthermore, \texttt{SERTAO} incorporates a hybrid interface designed to operate in conjunction with \texttt{CLASS} \cite{Blas:2011rf}. In this configuration, the background evolution computed within \texttt{SERTAO} can be consistently interfaced with the full Boltzmann treatment of perturbations provided by \texttt{CLASS}, enabling cross-validation and extended analyses. This hybrid capability allows us to maintain internal control over background modifications while preserving compatibility with state-of-the-art Boltzmann solvers for high-precision calculation.

Parameter inference is performed using the publicly available nested sampling algorithm \texttt{dynesty} \cite{Speagle:2019ivv}, which enables efficient exploration of multi-dimensional parameter spaces and provides robust estimates of posterior distributions as well as Bayesian evidence. This approach is particularly well suited for models with extended parameter spaces, where potential degeneracies and non-Gaussian posteriors may arise. Convergence and sampling reliability are monitored through the internal stopping criteria of the nested sampling algorithm, ensuring stability of the posterior reconstruction and evidence computation. The implementation allows seamless combination of multiple datasets, within a unified likelihood framework. This strategy guarantees numerical consistency between the theoretical model and the inferred cosmological observables, while maintaining full flexibility to incorporate generalized modifications such as deviations from the cosmic distance duality relation.

The \texttt{SERTAO} framework therefore constitutes a numerically stable, statistically robust, and computationally efficient platform for testing generalized cosmological scenarios, featuring a suite of fully implemented likelihoods that can be seamlessly combined within a unified inference pipeline.
In this work, we explored the parameter space constrains using \texttt{SERTAO}, ensuring convergence of all chains through the Gelman--Rubin diagnostic criterion~\cite{Gelman_1992}, requiring $R-1 \leq 10^{-2}$. 

Flat priors were adopted over the cosmological parameter set relevant to the datasets under consideration. In our analyses, we assume uniform priors for all baseline parameters with sufficiently wide ranges, namely
\begin{equation}
\begin{aligned}
p &\in [1.3, 3.0], \quad
m \in [0.0, 3.0], \quad
\Omega_b \in [0.02, 0.06], \\
\Omega_{\rm cdm} &\in [0.10, 0.50], \quad
H_0 \in [40, 90], \quad
\sigma_8 \in [0.5, 1.0]\,.
\end{aligned}
\end{equation}

The statistical outputs was carried out using the \texttt{GetDist} package\footnote{https://github.com/cmbant/getdist}, which enabled the extraction of numerical results, including one-dimensional posterior distributions and two-dimensional marginalized probability contours.

More detailed information about the datasets used follows below.

\begin{itemize}
    \item \textit{Growth measurements} (\textbf{Growth}): We consider the growth dataset compiled in Ref.~\cite{Avila:2022xad}, consisting of 20 $f\sigma_8(z)$ measurements spanning the redshift range $0.02<z<1.944$ and 11 $f(z)$ measurements over $0.013<z<1.40$. The former includes only direct measurements of $f\sigma_8$, while the latter contains only direct measurements of $f$, excluding $f\sigma_8$ measurements that rely on a fiducial cosmology to remove the dependence on $\sigma_8$. When multiple data releases are available for the same survey, only the most recent measurement is retained. Measurements from the same cosmological tracer are included only if they correspond to independent redshift bins, whereas correlated bins are considered only when different tracers are used. This conservative selection results in a smaller, but more statistically robust, dataset than some compilations adopted in the literature.
    
    \item \textit{Baryon Acoustic Oscillations} (\textbf{DESI-DR2}): Baryon Acoustic Oscillations (BAO) measurements provided by Dark Energy Spectroscopic Instrument (DESI) collaboration from observations of galaxies and quasars \cite{DESI:2024uvr}, and Lyman-$\alpha$ tracers \cite{DESI:2024lzq}, as summarized in Table I of Ref \cite{DESI:2024mwx}. These measurements consist of both isotropic
    and anisotropic BAO data in the redshift range $0.1 < z < 4.2$ and are divided into seven redshift bins. The isotropic BAO measurements are represented as DV(z)/rd, where DV denotes the angle-averaged distance, normalized to the (comoving) sound horizon at the drag epoch. The anisotropic BAO measurements include $DM(z)/rd$ and $DH(z)/rd$, where DM is the comoving angular diameter distance and $DH$ is the Hubble horizon. Additionally, the correlation between the measurements of $DM/rd$ and $DV/rd$ is also taken into account. We refer to this dataset as DESI-DR2.

    \item \textit{Type Ia Supernovae}: Type Ia supernovae act as standardizable candles, providing a crucial method for measuring the universe’s expansion history and supporting $\Lambda$-dominated models. In this work, we use the following recent samples: 
    \begin{enumerate}
        \item \textbf{PantheonPlus (PP)}: We incorporated SN Ia distance modulus measurements from the PantheonPlus sample \cite{Brout:2022vxf}, which consists of 1550 supernovae spanning a redshift range from $0.01$ to $2.26$. We refer to this dataset as PP.

        \item \textbf{Union 3.0 (U3)}: The Union 3.0 compilation, consisting of 2087 SN Ia, was presented in \cite{Brout:2022vxf}. Notably, 1363 of these SN Ia are common with the PantheonPlus sample. This dataset features a distinct treatment of systematic errors and uncertainties, employing Bayesian Hierarchical Modeling. We refer to this dataset as U3.
    \end{enumerate}
\end{itemize}

Throughout this work, we include state-of-the-art Big Bang Nucleosynthesis (BBN) constraints on the physical baryon density, $\omega_{\rm b}=\Omega_{\rm b}h^2$, in all dataset combinations. The \textbf{BBN} likelihood is based on the primordial helium mass fraction, $Y_{\rm P}$, from Ref.~\cite{Aver_2015}, and the primordial deuterium abundance, $(\mathrm{D}/\mathrm{H})_{\rm P}$, from Ref.~\cite{Cooke_2018}.

\section{Observational constraints from growth-rate measurements}
\label{sec:constrains}

First, considering the simplest case with $m=1$, we found that the growth data already provide strong constraining capacity, despite the additional freedom introduced by the model. Even so, the inclusion of BAO and supernova data yields a notable further improvement in the parameter constraints, significantly reducing the posterior uncertainties across all parameters, as summarized in Table~\ref{tab:constraintsI}.

\begin{table*}[htpb!]
\centering
\footnotesize
\renewcommand{\arraystretch}{1.5}
\caption{Marginalized constraints and mean values with 68\% CL for the parameters of MOND-like framework, assuming a fixed $m=1$. In the last rows, we present the quantity $\Delta \chi _{\rm min} ^2 = \Delta \chi _{\rm MOND-like} ^2 - \Delta \chi _{\Lambda CDM} ^2$, which compare the model fits. Negative value for it indicate a preference for this model over the $\Lambda$CDM model, while positive values favor the $\Lambda$CDM model.}
\label{tab:constraintsI}
\resizebox{\textwidth}{!}{
\begin{tabular}{lcccc}
\hline\hline
\textbf{Parameter} & \textbf{Growth} & \textbf{Growth+DESI-DR2} & \textbf{Growth+DESI-DR2+PP} & \textbf{Growth+DESI-DR2+U3} \\ 
\hline\hline
$p$ & $2.08\pm 0.13$  & $2.080\pm 0.088$ & $2.109\pm 0.086$ & $2.109^{+0.079}_{-0.091}$\\
$\Omega_{\mathrm{m}}$  & $0.306^{+0.045}_{-0.058}$  & $0.296\pm 0.014$ & $0.311\pm 0.012$ & $0.311\pm 0.013$ \\
$S_8$ & $0.835^{+0.083}_{-0.10}$ & $0.822^{+0.043}_{-0.049}$ & $0.846^{+0.042}_{-0.048}$ & $0.846^{+0.043}_{-0.048}$ \\ \hline
$\Delta \chi_{\rm min}^2 $ & $-0.25$ & $-0.59$ & $-1.27$ & $-1.25$ \\
\hline\hline
\end{tabular}}
\end{table*}

In all dataset combinations, the parameter $p$ is tightly constrained around $p \simeq 2.1$, indicating a mild but persistent deviation from the standard gravitational expectation preferred by the data. The matter density parameter $\Omega_m$ remains fully consistent with $\Lambda$CDM values, showing no significant shift across different combinations. For $S_8$, the model does not alleviate the well-known tension, instead, it slightly favors a higher clustering amplitude while preserving internal consistency among the datasets. Overall, the negative values of $\Delta \chi^2_{\rm min}$ across all cases in Table~\ref{tab:constraintsI} indicate a mild but robust preference for the MOND-like extension over $\Lambda$CDM in the LSS sector alone. Although not statistically significant, this trend suggests that the additional degree of freedom introduced by $p$ provides a marginal improvement in the fit, while remaining fully consistent with background cosmological constraints.

\begin{figure}[htpb!]
    \centering
    \includegraphics[width=0.7\columnwidth]{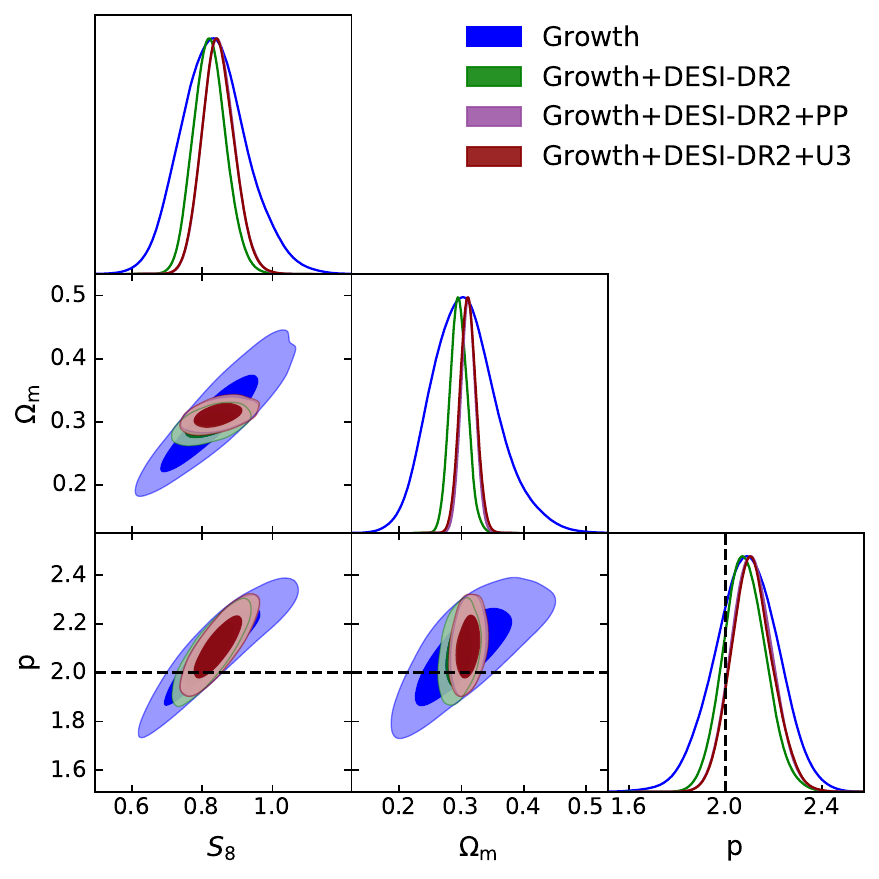} 
    \caption{Posterior distributions 68\% and 95\% credible contours for the parameters $S_{8}$, $\Omega_{\rm m}$, and $p$.}
    \label{fig:triangleplot}
\end{figure}

In Figure~\ref{fig:triangleplot} we observe the parameter correlations introduced by the MOND-like modification (for $m=1$). We first note a strong positive correlation between $p$ and $S_8$, which remains robust across all dataset combinations considered. The origin of this correlation can be understood directly from the modified growth equation \eqref{eq:growth}. The parameter $p$ enters through the effective gravitational coupling $\mu_{\rm eff}$, modifying the source term responsible for the growth of matter perturbations. In the cosmologically relevant regime, we find that larger values of $p$ generally increase the effective gravitational strength, leading to more efficient structure growth and, consequently, to larger values of $S_8$. A positive correlation is also found between $p$ and $\Omega_{\rm m}$, which reflects the fact that both parameters contribute to enhancing the growth of matter perturbations.

Table~\ref{tab:constraintsII} summarizes the results when both $p$ and $m$ vary. Across all data combinations, the best-fit values of $p$ lie above $2$, reaching $p=2.162^{+0.057}_{-0.18}$ for the full Growth+DESI-DR2+U3 compilation. This suggests a mild phenomenological preference for a suppressed effective gravitational coupling, $\mu{\rm eff}<1$. However, the $\Lambda$CDM limit, $p=2$, is consistent with the results at the $1\sigma$ level in every case. 

\begin{table*}[htpb!]
\centering
\footnotesize
\renewcommand{\arraystretch}{1.5}
\caption{Same as table \ref{tab:constraintsI}, but assuming $m$ as a free parameter.}
\label{tab:constraintsII}
\resizebox{\textwidth}{!}{
\begin{tabular}{lcccc}
\hline\hline
\textbf{Parameter} & \textbf{Growth} & \textbf{Growth+DESI-DR2} & \textbf{Growth+DESI-DR2+PP} & \textbf{Growth+DESI-DR2+U3}\\
\hline\hline 
$p$ & $2.13^{+0.14}_{-0.23}$ & $2.117^{+0.070}_{-0.16}$ & $2.170^{+0.062}_{-0.19}$ & $2.162^{+0.057}_{-0.18}$\\
$m$ & $0.98^{+0.60}_{-0.99}$ & $1.03^{+0.37}_{-1.0}$ & $0.999^{+0.68}_{-1.0}$& $1.04^{+0.36}_{-1.0}$ \\
$\Omega_m$  &  $0.306^{+0.048}_{-0.060}$ & $0.296^{+0.013}_{-0.015}$ & $0.311\pm 0.012$ & $0.311^{+0.012}_{-0.014}$\\
$S_8$  & $0.835\pm 0.096$ & $0.824^{+0.043}_{-0.050}$ & $0.846^{+0.042}_{-0.047}$ & $0.846^{+0.043}_{-0.048}$\\ \hline
$\Delta \chi_{\rm min}^2 $ & $-0.25$ & $-0.55$ & $-1.41$ & $-1.21$ \\
\hline\hline
\end{tabular}}
\end{table*}

\begin{figure}[htpb!]
    \centering
    \includegraphics[width=\columnwidth]{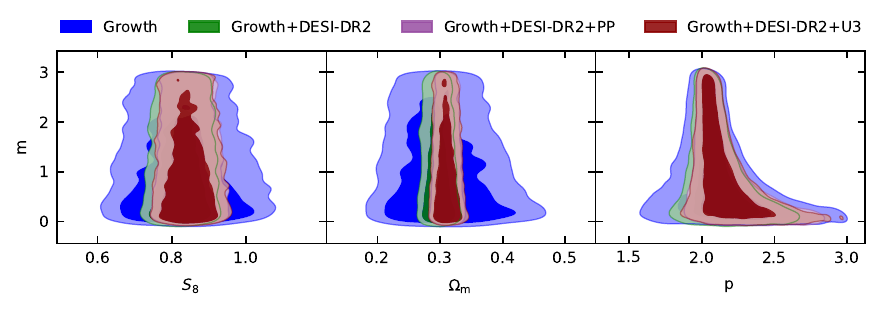} 
    \caption{Constraints at the $68\%$ and $95\%$ confidence levels for $S_8$, $\Omega_{\rm m}$, and $p$, with a free $m$. The different colors denote the dataset combinations listed in the legend.}
    \label{fig:retagleplot}
\end{figure}

In Fig.~\ref{fig:retagleplot} we find an overall absence of clear correlations among parameters when $m$ is allowed to vary freely, reflecting a strong degeneracy structure in the extended parameter space. In this regime, the additional freedom associated with $m$ weakens the constraining power of the data, leading to poorly defined correlation patterns. More specifically, the effects of the modified gravitational dynamics encoded in $m$ can be largely absorbed by correlated shifts in $p$, $\Omega_m$, and $S_8$. This degeneracy implies that changes in $m$ do not produce independent observational signatures, but are instead reabsorbed by the other parameters of the model.

\begin{figure}[htpb!]
    \centering
    \includegraphics[width=\columnwidth]{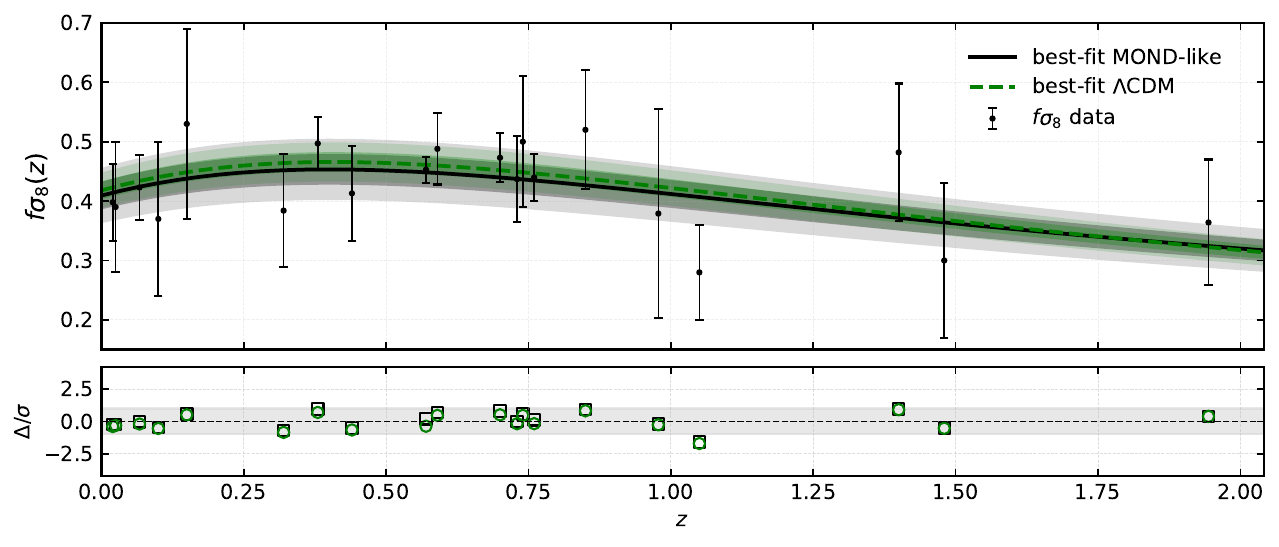} 
    \caption{Statistical reconstruction of the theoretical $f\sigma_8(z)$ prediction at the $1\sigma$ and $2\sigma$ confidence levels for the $\Lambda$CDM and MOND-like models (with $m$ fixed), obtained from the joint \textit{Growth+DESI-DR2+PP} analysis. Black points denote the $f\sigma_8$ measurements compiled in Ref.~\cite{Avila:2022xad}. The lower panel displays the residuals between the model predictions and the $f\sigma_8(z)$ measurements, normalized by the corresponding observational uncertainties.}
    \label{fig:fsigma8}
\end{figure}

Overall, our comprehensive analysis reinforces the importance of growth rate measurements in providing observational constraints on MOND-like modifications to gravity. To gain deeper insight into the origin of these constraints, Figure ~\ref{fig:fsigma8} presents the theoretical predictions of $f\sigma_8 (z)$ for both the MOND-like models and $\Lambda$CDM. As in Figure \ref{fig:f_z}, higher values of $p$ decrease the theoretical prediction of $f\sigma_8 (z)$, suppressing the overall amplitude of the curve, whereas lower values of $p$ increase it. Notably, the largest discrepancy between the two models occurs at low redshifts, underscoring the importance of late-time data in distinguishing between the models and constraining the associated cosmological parameters.

\section{Modeling Corrections to Matter Power Spectrum Predictions}
\label{sec:eft}

The non-linear regime of structure formation offers powerful constraints on cosmological parameters beyond the linear approximation. In this regime, gravitational evolution becomes intrinsically nonlinear, and different Fourier modes no longer evolve independently but instead couple to one another, leading to a transfer of power across scales and the generation of non-Gaussian features in the matter distribution. Physically, this corresponds to the formation of collapsed and virialized structures such as dark matter halos, galaxies, and clusters. Reinterpreting the non-linear regime through the Effective Field Theory of Large-Scale Structure (EFTofLSS) \cite{Baumann:2010tm,Carrasco:2012cv, Carrasco:2013mua}(see, e.g., \cite{Ivanov:2022mrd, Ivanov:2026nlf} for a comprehensive recent review) provides a consistent framework. In contrast with SPT, the EFTofLSS introduces an effective stress tensor for the matter fluid that accounts for
imperfect fluid corrections, such as viscosity and velocity dispersion. The terms in the effective stress tensor
contain all operators allowed by the equivalence principle, where the non-linear evolution of matter and the small-scale cutoff dependence are naturally absorbed into counterterms. These free parameters are subsequently evaluated using observational datasets. 

The modeling of non-linear scales within the one-loop EFTofLSS framework has already been successfully implemented across numerous galaxy clustering full-shape analyses, becoming well-established in the literature \cite{Philcox:2021kcw, DESI:2024hhd, fR1, Silva:2025twg, Chudaykin:2026nls, Chudaykin:2020aoj}. Furthermore, the formalism has been confronted with other cosmological probes, such as the effective field theory of Lyman-$\alpha$  \cite{Ivanov:2023yla, Karacayli:2026wtw, deBelsunce:2025gci, He:2025jwp} and weak lensing cosmic shear studies \cite{Chen:2026usz, Saraivanov:2026sxc, DAmico:2025zui}. In particular, recent weak lensing applications in the two-loop order \cite{Saraivanov:2026sxc, Chen:2026usz} have yielded comparable, or even superior, constraining power on cosmological parameters when employing the EFTofLSS instead of simulation-based matter power spectra. Collectively, these works demonstrate the efficacy of this framework in parametrizing small-scale physics when applied to real data. Moreover, recent advances toward a consistent framework for the two-loop galaxy power spectrum within EFTofLSS \cite{Ivanov:2026zos, Bakx:2025jwa, Bakx:2026rmd} highlight a promising future for the field, opening the way to push our understanding of structure formation through the non-linear regime.

%However, as upcoming large-scale structure surveys push observational uncertainties to the sub-percent level, an accurate robust treatment of this regime becomes indispensable to avoid theoretical biases and accurately constrain alternative gravity models. Under these considerations, it is crucial to advance beyond linear order to compute the higher-order loop contributions in the structure formation regime. In this work, we focus on extending this non-linear formalism to non-standard $\Lambda$CDM frameworks.

This formalism has also been extended to other cosmologies besides $\Lambda$CDM, we leave here just a few refereences \cite{Silva:2025bnn, Verhoeve:2026gtt, Pal:2025zep, Ishak:2024jhs, Bottaro:2023wkd}. In this section, we compute the full one-loop matter power spectrum in a MOND-like class of models, starting by evaluating the perturbative kernels and the corresponding loop integrals. Unlike the standard cosmological framework, the modified gravitational dynamics in our specific scenario inherently introduce an extra scale and time dependence into the governing fluid equations via the effective coupling $\mu_{\rm eff}(k, \tau)$.

\begin{figure}[htpb!]
    \centering
    \includegraphics[scale=0.44]{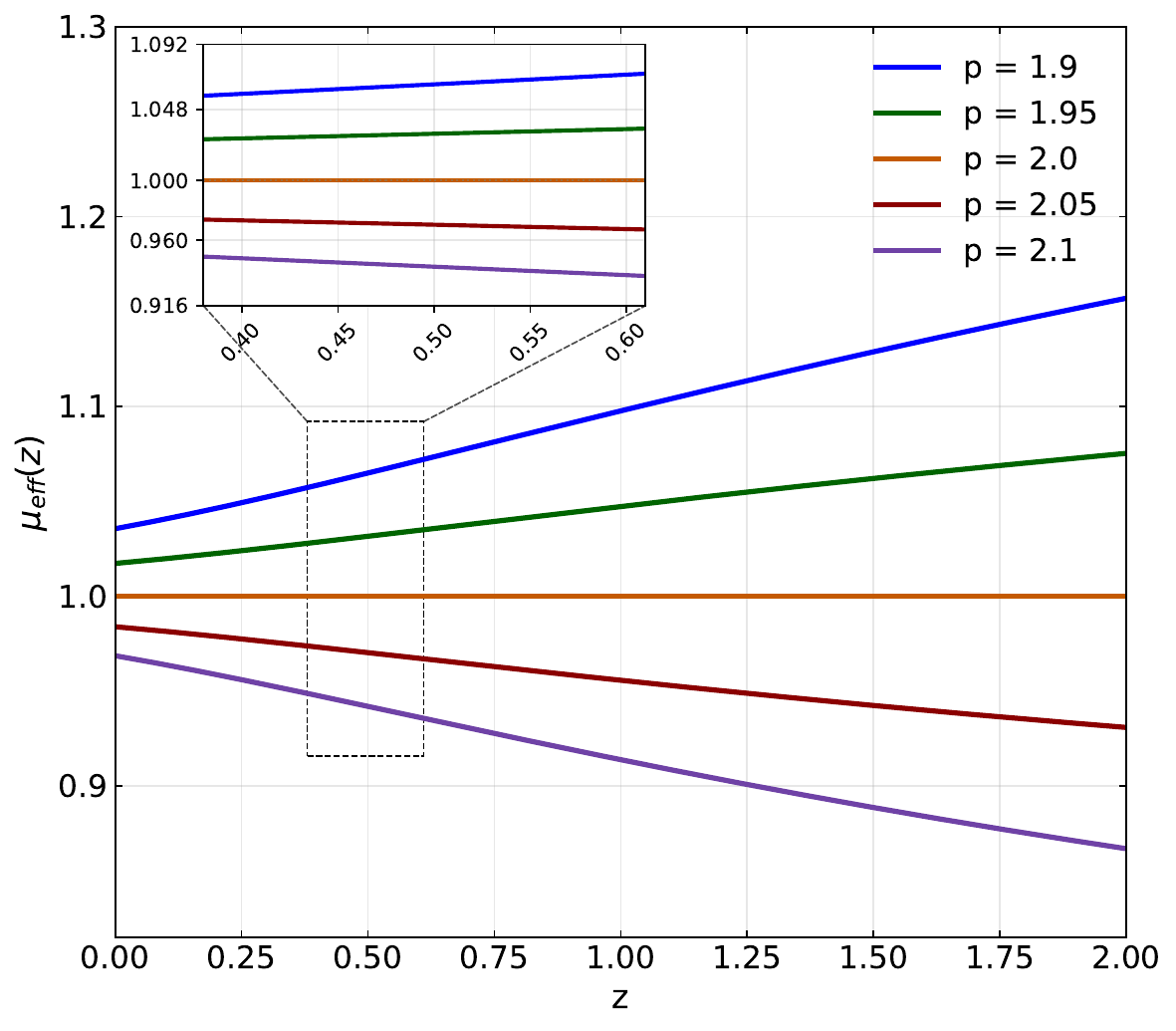}
    \includegraphics[scale=0.44]{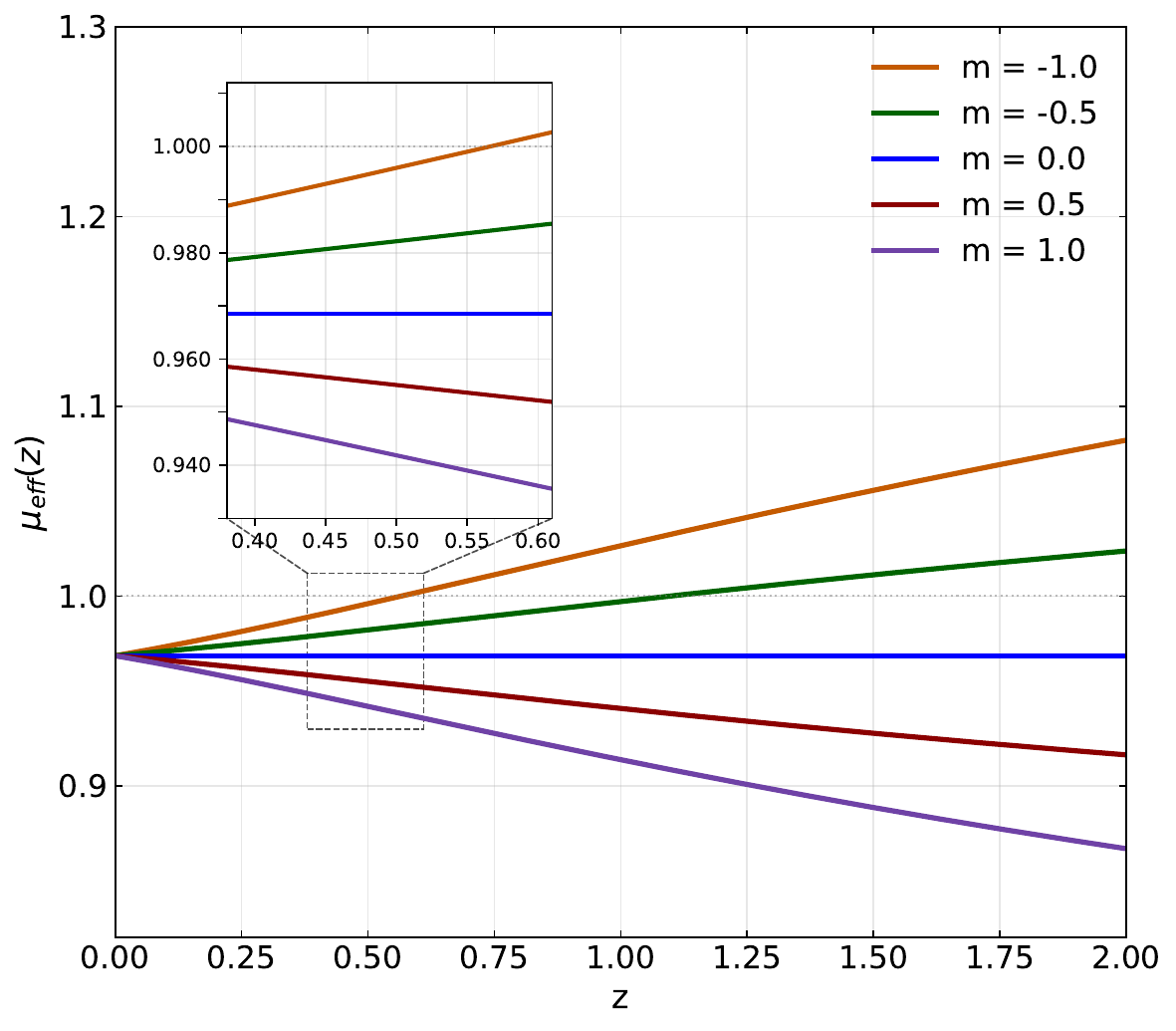}
    \caption{\textit{Left panel:} Redshift evolution of the effective gravitational coupling $\mu_{\rm eff}(\mathbf{k}, \mu, \tau)$ in our specific MOND-like scenario. The left panels illustrate the dependence on the parameter $p$ (fixing $m=1$), while the right panels show the dependence on the parameter $m$ (fixing $p=2.10$, from the best-fit). We zoom on the BOSS DR12 observational window ($z \in [0.38, 0.61]$. Within this narrow observational bin, $\mu_{\rm eff}(z)$ varies by less than $\sim 2\%$, justifying the use of the quasi-static approximation ($\mu_{\rm eff}(\tau) \approx \text{const.}$) for the spatial momentum loop integrations in the main text.}
    \label{fig:mueff}
\end{figure}

Nevertheless, Fig.~\ref{fig:mueff} shows the time evolution of this additional term over the full redshift range covered by the growth data, with a zoom-in highlighting the range probed by some current galaxy clustering full-shape analyses. Within these restricted ranges, the extra dynamical contribution evolves sufficiently slowly that it can be effectively decoupled from the spatial integrations without any significant loss of predictive power. In order to maintain the clarity of the main text, the general argument and the rigorous derivation of the full time-dependent solutions are deferred to subsection (\ref{sec:appA}). Consequently, throughout this section, we assume that $\mu_{\rm eff}$ depends solely on $k$ and $\mu$.

We start by using the Fourier definition (\ref{fourier}) on the continuity equation (\ref{eq:cont}) \footnote{ 
The integral is defined as:
\begin{equation}
    \int \left(\prod_{i=1}^{n}\frac{d^3q_i}{(2\pi)^3}\right) (2\pi)³ \ [...]  = \int _{q_1 \ \cdots \ \ q_n}  [...] \ .
\end{equation} }:
\begin{equation}
\frac{\partial \delta(\mathbf{k}, \tau)}{\partial \tau} + \theta(\mathbf{k}, \tau) = 
- \int _{q_1, q_2} \delta_D(\mathbf{k} - \mathbf{q}_1 - \mathbf{q}_2) \, \alpha(\mathbf{q}_1, \mathbf{q}_2) \, \theta(\mathbf{q}_1, \tau) \, \delta(\mathbf{q}_2, \tau). 
\end{equation}
Now we can take the divergent of Euler equation (\ref{eq:euler}) and rewrite it in real space as:
\begin{align}
\frac{\partial \theta(\mathbf{x}, \tau)}{\partial \tau} + \mathcal{H} \theta(\mathbf{x}, \tau) + \nabla \cdot \left[(\mathbf{u}(\mathbf{x}, \tau) \cdot \nabla) \mathbf{u}(\mathbf{x}, \tau)\right] &= - \nabla^2 \Phi(\mathbf{x}, \tau) = - \nabla^2 \varphi(\mathbf{x}, \tau),
\end{align}
this equation in Fourier space has the form:
\begin{equation}
    \frac{\partial \theta(\mathbf{k}, \tau)}{\partial \tau} 
+ \mathcal{H}(\tau) \theta(\mathbf{k}, \tau) 
+ \frac{3}{2} \Omega_m \mathcal{H}^2(\tau) \mu_{\rm eff}(\mathbf{k},\mu)\delta(\mathbf{k}, \tau) 
= \nonumber \\
- \int _{q_1q_2} \delta_D(\mathbf{k} - \mathbf{q}_1 - \mathbf{q}_2) 
\, \beta(\mathbf{q}_1, \mathbf{q}_2) \theta(\mathbf{q}_1, \tau)\theta(\mathbf{q}_2, \tau).  
\end{equation}
In these equations, $\alpha$ and $\beta$ are the standard mode coupling, that arises naturally from the nonlinear terms in the equations. Physically, this coupling reflects the fact that, beyond the linear regime, modes with different wavevectors $\mathbf{q}_1$ and $\mathbf{q}_2$ no longer evolve independently but instead interact and exchange power.
\begin{equation} 
\alpha(\mathbf{q}_1, \mathbf{q}_2) = 1 + \frac{\mathbf{q}_1 \cdot \mathbf{q}_2}{q_1^2},
\end{equation}
and
\begin{equation}
    \beta(\mathbf{q}_1, \mathbf{q}_2) = \frac{(\mathbf{q}_1 \cdot \mathbf{q}_2)(q_1 + q_2)^{2}}{2 q_1^2 q_2^2}.
\end{equation}

Without the extra time dependence, we can retain the same perturbative structure commonly used in $\Lambda$CDM and to employ the so-called Einstein–de Sitter (EdS) approximation \cite{eds1, eds2, eds3} allowing the perturbative expansion to preserve the standard kernel structure while incorporating the modified gravitational effects. With this choice the non linear fields can be expanded as: 
\begin{align}
\delta(\mathbf{k}, \tau) &= \sum_{n=1}^{\infty} a^n(\tau) \, \delta_n(\mathbf{k}), \\\theta(\mathbf{k}, \tau) &= -\mathcal{H}(\tau) \sum_{n=1}^{\infty} a^n(\tau) \,\theta_n(\mathbf{k}),
\end{align}
The perturbative framework of the SPT allows us to express the higher-order density solutions, $\delta_n$, as convolution integrals of the linear density field \cite{Bernardeau:2001qr}:
\begin{equation}
\delta_n(\mathbf{k}) = \int _{q_1  \cdots  q_n }\, \delta_D\Bigg( \mathbf{k} - \sum_{i=1}^n \mathbf{q}_i \Bigg) 
F_n(\mathbf{q}_1, \ldots, \mathbf{q}_n) \, \delta_1(\mathbf{q}_1) \cdots \delta_1(\mathbf{q}_n),
\end{equation}
where \(F_n\) are the perturbative kernels at order \(n\), and \(\delta_1(\mathbf{q})\) is the linear density contrast in Fourier space. In particular, we are interested in computing the contributions to the matter power spectrum arising from the second- and third-order terms, \(\delta_2(\mathbf{k})\) and \(\delta_3(\mathbf{k})\), whose Fourier-space expressions are:

\begin{equation}
\delta_2(\mathbf{k}) = \int _{q_1 q_2} \delta_D(\mathbf{k} - \mathbf{q}_1 - \mathbf{q}_2) 
F_2(\mathbf{q}_1, \mathbf{q}_2) \, \delta_1(\mathbf{q}_1) \, \delta_1(\mathbf{q}_2),
\end{equation}
and
\begin{equation}
\delta_3(\mathbf{k}) = \int _{q_1 q_2 q_3} \delta_D(\mathbf{k} - \mathbf{q}_1 - \mathbf{q}_2 - \mathbf{q}_3) 
F_3(\mathbf{q}_1, \mathbf{q}_2, \mathbf{q}_3) \, \delta_1(\mathbf{q}_1) \, \delta_1(\mathbf{q}_2) \, \delta_1(\mathbf{q}_3).
\end{equation}

In this work, however, the perturbation kernels depart from the standard Einstein-de Sitter templates. Instead, they are substituted by a new formulation arising from the MOND-like framework. Solving the underlying governing equations perturbatively reveals that the recursive relations for these kernels exhibit a distinct deviation from the standard case:

\begin{align}
F_n^{\rm Mond}(\mathbf{q}_1, \ldots, \mathbf{q}_n) &=
\sum_{m=1}^{n-1} 
\frac{G_m^{\rm Mond}(\mathbf{q}_1, \ldots, \mathbf{q}_m)}{n(2n +1) -3\mu_{\rm eff}}
\Big[
(2n+1)\,\alpha(\mathbf{k}_1, \mathbf{k}_2)
F_{n-m}^{\rm Mond}(\mathbf{q}_{m+1}, \ldots, \mathbf{q}_n)
\nonumber \\
&\quad + 2\,\beta(\mathbf{k}_1, \mathbf{k}_2)
G_{n-m}^{\rm Mond}(\mathbf{q}_{m+1}, \ldots, \mathbf{q}_n)
\Big],
\\[1em]
G_n^{\rm Mond}(\mathbf{q}_1, \ldots, \mathbf{q}_n) &=
\sum_{m=1}^{n-1} 
\frac{G_m^{\rm Mond}(\mathbf{q}_1, \ldots, \mathbf{q}_m)}{n(2n+1) -3\mu_{\rm eff}}
\Big[
3\,\mu_{\rm eff}\alpha(\mathbf{k}_1, \mathbf{k}_2)
F_{n-m}^{\rm Mond}(\mathbf{q}_{m+1}, \ldots, \mathbf{q}_n)
\nonumber \\
&\quad + 2n\,\beta(\mathbf{k}_1, \mathbf{k}_2)
G_{n-m}^{\rm Mond}(\mathbf{q}_{m+1}, \ldots, \mathbf{q}_n)
\Big],
\end{align}

\noindent
where $\mathbf{k}_1 \equiv \mathbf{q}_1 + \ldots + \mathbf{q}_m$, 
$\mathbf{k}_2 \equiv \mathbf{q}_{m+1} + \ldots + \mathbf{q}_n$, 
$\mathbf{k} \equiv \mathbf{k}_1 + \mathbf{k}_2$, $F_1 = G_1 = 1$, and we have defined the MOND-like kernels $F_n^{\rm Mond}$ and $G_n^{\rm Mond}$. Clearlly, when $\mu_{\rm eff} = 1$ we take back $\Lambda$CDM.
Additionally, it is very useful to express the perturbative kernels in our case in terms of the standard $\Lambda$CDM kernels. 

By exploiting the recursive relations that define the hierarchy of SPT kernels, we can reorganize the modified solutions so that they are written as corrections to the usual $F_n^{\rm std}$ and $G_n^{\rm std}$ $\Lambda$CDM kernels. Using these recursive relations, we find that the kernels that appears in the one-loop contribuiton are:

\begin{align}
\label{F_2}
F_2^{\rm Mond}(\mathbf{q}_1,\mathbf{q}_2) 
&= \frac{7}{10 - 3\mu_{\rm eff}}F_2^{\rm std}(\mathbf{q}_1,\mathbf{q}_2) , \\[6pt]
G_2^{\rm Mond}(\mathbf{q}_1,\mathbf{q}_2) 
&= G_2^{\rm std}(\mathbf{q}_1,\mathbf{q}_2) + \lambda(\mathbf{q}_1,\mathbf{q}_2) , \ 
\end{align}

and the third order kernel can be written as:

\begin{align}
\label{F_3}
F_3^{\mathrm{Mond}}(\mathbf{q}_1,\mathbf{q}_2,\mathbf{q}_3) = & \, \frac{18}{21-3\mu_{\mathrm{eff}}} F_3^{\mathrm{std}}(\mathbf{q}_1,\mathbf{q}_2,\mathbf{q}_3) \nonumber \\
& + \frac{1}{21-3\mu_{\mathrm{eff}}} \Biggl[ 7\alpha(\mathbf{q}_1,\mathbf{q}_{23})\Delta(\mathbf{q}_2,\mathbf{q}_3) + 2\beta(\mathbf{q}_1,\mathbf{q}_{23})\lambda(\mathbf{q}_2,\mathbf{q}_3) \nonumber \\
& + 7\lambda(\mathbf{q}_1,\mathbf{q}_2)\alpha(\mathbf{q}_{12},\mathbf{q}_3) + 2\lambda(\mathbf{q}_1,\mathbf{q}_2)\beta(\mathbf{q}_{12},\mathbf{q}_3) \Biggr],
\end{align}
where $ \mathbf{q_{ij} = q_i + q_j}$, and we have defined the following function of the wave vectors:
\begin{align}
\Delta(\mathbf{q_1,q_2}) & =
\frac{15\alpha(\mathbf{q_1,q_2}) + 6\beta(\mathbf{q_1,q_2}) }{7(10-3\mu_{\rm eff})}(\mu_{\rm eff} - 1) \\
\lambda(\mathbf{q_1,q_2}) &=
\frac{30\alpha(\mathbf{q_1,q_2}) + 12\beta(\mathbf{q_1,q_2}) }{7(10-3\mu_{\rm eff})}(\mu_{\rm eff} - 1).
\end{align}

These additional functions encode corrections to the standard $\Lambda$CDM dynamics. In particular, they parameterize deviations induced by the modified gravitational sector and therefore represent departures from the standard perturbative kernels. By construction, all these contributions vanish in the limit $\mu_{\rm eff} = 1$, in which case the standard $\Lambda$CDM expressions are fully recovered.

Furthermore, it is convenient to introduce the symmetrized kernels, that appears in the one-loop corrections, constructed by summing over all distinct permutations of their arguments:
\begin{equation}
\label{Fs}
    F_{\space n} ^{(s)} = \sum_n \frac{1}{n!} [F(\mathbf{q_1, \cdots, q_n}) + \cdots +F(\mathbf{q_n}, \cdots, \mathbf{q_1})]
\end{equation}

At this point, we investigate how the modifications to the matter density contrast propagate into the matter power spectrum within the MOND-like framework. In particular, we assess how the  parametrization introduced in equation (\ref{GPE}) alters the evolution of perturbations and, consequently, the resulting shape of $P(k)$ when compared to the standard $\Lambda$CDM case.

We can define the correlations functions that appears on the 1-loop power spectrum expansion as:
\begin{equation}
\langle \delta_2(\mathbf{k}) \delta_2(\mathbf{k}') \rangle = (2\pi)^3 \delta_D(\mathbf{k} + \mathbf{k}') P_{22}(k), \qquad \langle \delta_{1}(\mathbf{k}) \delta_3(\mathbf{k}') \rangle = (2\pi)^3 \delta_D(\mathbf{k} + \mathbf{k}') P_{13}(k).
\end{equation}

For the first one-loop contribution, $P_{22}(k)$, performing the required algebraic manipulations yields
\begin{equation}\label{P22new}
P_{22}^{\rm  Mond}(k) = 
2 \int d^3 q \,
\left[ F_2^{\rm Mond \space (s)}(\mathbf{q}, \mathbf{k}-\mathbf{q}) \right]^2
P_L(|\mathbf{k}-\mathbf{q}|)\, P_L(q),
\end{equation}
where $F_2^{\rm Mond \space (s)} $ stands for the symmetrical Mond-like $F_2$ kernel. By using equation (\ref{F_2}) it is easy to see that:
\begin{equation}
\label{eq:P22mond}
    P_{22}^{\rm Mond} = \left(\frac{7}{10 - 3\mu_{\rm eff}} \right)^2 P_{22}^{\rm std}.
\end{equation}

This equation shows how the $P_{22}$ term appearing in the one-loop correction is affected by the MOND-like scenario. We emphasize that the notation $P_{22}^{\rm std}$ refers to the $\Lambda$CDM case (SPT framework). As expected, in the limit $\mu_{\rm eff} = 1$ the standard result is fully recovered.

A similar procedure can be applied to the \(P_{13}(k)\) contribution. where:
\begin{equation}
\label{eq:P13mond}
    P_{13}^{\rm Mond}(k) = 6\, P_L(k) \int d^3q\, F_3^{\rm Mond \space (s)}(\mathbf{k}, \mathbf{q}, -\mathbf{q})\, P_L(q) ,
\end{equation}
so by using the result of equation (\ref{F_3}) we can rewrite que $P_{13}$ contribution as:
\begin{equation}
\label{eq:P13new}
P_{13}^{\rm Mond}(k) = \frac{18}{21-3\mu_{\rm eff}}P_{13}^{\rm std}(k) + \Delta P_{13}(k),
\end{equation}
where $\Delta P_{13}$ stands for:

\begin{equation}
\begin{aligned}
\label{DP13}
\Delta P_{13} =\frac{6P_L(k)}{21-3\mu_{\rm eff}} \int d^3\mathbf{q} \, \Biggl[ & 7\alpha(\mathbf{q}_1,\mathbf{q}_{23})\Delta(\mathbf{q}_2,\mathbf{q}_3) + 2\beta(\mathbf{q}_1,\mathbf{q}_{23})\lambda(\mathbf{q}_2,\mathbf{q}_3) \\[1ex]
& + 7\lambda(\mathbf{q}_1,\mathbf{q}_2)\alpha(\mathbf{q}_{12},\mathbf{q}_3) + 2\lambda(\mathbf{q}_1,\mathbf{q}_2)\beta(\mathbf{q}_{12},\mathbf{q}_3) \Biggr] .
\end{aligned}
\end{equation}
Here, we use the notation \(\Delta P_{13}(k)\) to quantify the total deviation from the standard case. We emphasize that the integrand in equation (\ref{DP13}) requires full symmetrization with respect to its arguments, following the structure presented in equation (\ref{Fs}). Given that the complete explicit expression is extremely lengthy, it is omitted here for brevity. In addition, after simetrizing, we explicitly substitute the wave vectors associated with the $P_{13}$ contribution\footnote{We set  
\(\mathbf{q}_1 = \mathbf{k}\), \(\mathbf{q}_2 = \mathbf{q}\), \(\mathbf{q}_3 = -\mathbf{q}\),  
\(\mathbf{q}_{23} = \mathbf{q}_2 + \mathbf{q}_3 = \mathbf{0}\), and \(\mathbf{q}_{12} = \mathbf{q}_1 + \mathbf{q}_2 = \mathbf{k} + \mathbf{q}\).} into the integrand of equation (\ref{DP13}).

However, numerically evaluating the loop integrals in Eq.~\eqref{P22new} and ~\eqref{eq:P13mond} presents a significant computational challenge. These integrals involve multi-dimensional convolutions that are slow to calculate using standard numerical integration. A clever solution to this problem is the FFTLog algorithm. This method computes a Fast Fourier Transform of a function over logarithmically spaced intervals~\cite{Hamilton:1999uv}. In cosmology, its main advantage developed in \cite{Simonovic:2017mhp} is that it transforms complex loop integrals into simple matrix multiplications. This allows the calculation to be performed very quickly using standard linear algebra routines. Consequently, this approach is widely used in EFTofLSS codes, such as \textsc{CLASS-PT}\cite{Chudaykin:2020aoj}, \textsc{PyBird}\cite{DAmico:2020kxu}, and \textsc{FolpsD}\cite{DESI:2026haa}.

The formalism consists of decomposing the approximated linear real-space matter power spectrum into a series of power laws:
\begin{equation}
\bar{P}_{\rm L} (k) = \sum_{m = -N/2}^{N/2} c_m , k^{\nu + i\eta_m}_n ,
\end{equation}
where $N$ is the number of points sampled logarithmically over the $k$-interval. The Fourier coefficients $c_m$ and the discrete frequencies $\eta_m$ are given by
\begin{equation}
c_m = \frac{1}{N} \sum_{j=0}^{N-1} P_{\rm L}(k_j)k_j^{-\nu} k_{\rm min}^{-i\eta_m} e^{-2\pi i m j/N}, \qquad
\eta_m = \frac{2\pi m}{\ln(k_{\rm max}/k_{\rm min})}  .
\end{equation}
The parameter $\nu$ is an arbitrary real number known as the FFTLog bias..

At this point, we can apply this decomposition to the loop integrals, starting with the $P_{22}(k)$ term in Eq.~\eqref{eq:P22mond}. Substituting the power-law expansion into the integral yields
\begin{equation}
\label{eq:fftlog22}
    \bar{P}_{22}(k) = 2 \sum_{m_1,m_2} c_{m_1} c_{m_2} \sum_{n_1,n_2} f_{22}(n_1,n_2)\, k^{-2(n_1+n_2)} 
\int_{\mathbf{q}} \frac{1}{q^{2\nu_1 - 2n_1} |\mathbf{k}-\mathbf{q}|^{2\nu_2 - 2n_2}} \, ,
\end{equation}
where $n_1$ and $n_2$ are integer powers of $q^2$ and $|\mathbf{k}-\mathbf{q}|^2$ arising from the expansion of $F_2^2(\mathbf{q}, \mathbf{k}-\mathbf{q})$, and $f_{22}(n_1, n_2)$ denote the corresponding expansion coefficients. Here, we have defined the complex exponents
\begin{equation}
    \nu_1 = -\tfrac{1}{2}(\nu + i \eta_{m_1}) \quad \text{and} \quad \nu_2 = -\tfrac{1}{2}(\nu + i \eta_{m_2}) \, .
\end{equation}

A key advantage of this formalism is that it drastically reduces the computational cost by admitting an analytical solution for the momentum integral in Eq.~\eqref{eq:fftlog22}:
\begin{equation}
    \int_{\mathbf{q}} \frac{1}{q^{2\nu_1 - 2n_1} |\mathbf{k}-\mathbf{q}|^{2\nu_2 - 2n_2}} = k^{3 - 2(\nu_{12} - n_{12})}\, I(\nu_1 - n_1, \nu_2 - n_2) \, ,
\end{equation}
where $\nu_{12} \equiv \nu_1 + \nu_2$, $n_{12} \equiv n_1 + n_2$, and the dimensionless master integral $I(\nu_1, \nu_2)$ is defined as
\begin{equation}
    I(\nu_1, \nu_2) = \frac{1}{8\pi^{3/2}} \frac{\Gamma\left(\frac{3}{2}-\nu_1\right) \Gamma\left(\frac{3}{2}-\nu_2\right) \Gamma\left(\nu_{12}-\frac{3}{2}\right)}{\Gamma(\nu_1)\Gamma(\nu_2)\Gamma(3-\nu_{12})} \, .
\end{equation}

Furthermore, $I(\nu_1, \nu_2)$ satisfies well-known recurrence relations \cite{Simonovic:2017mhp}, which allow shifted terms such as $I(\nu_1+1, \nu_2)$ to be expressed as algebraic factors times $I(\nu_1, \nu_2)$ itself. Summing over all shifted terms enables us to factorize the entire expression into a matrix multiplication:
\begin{equation}
    \bar{P}_{22}(k) = k^3 \sum_{m_1,m_2} c_{m_1} k^{-2\nu_1} \cdot M_{22}(\nu_1,\nu_2) \cdot c_{m_2} k^{-2\nu_2} \, .
\end{equation}
In our framework, the matrix $M_{22}(\nu_1, \nu_2)$ is modified due to the presence of the generalized $F_2$ kernel in Eq.~(\ref{F_2}). Explicitly, it takes the form:
\begin{equation}
\begin{aligned}
M_{22}(\nu_1, \nu_2) = {} & \frac{(-3 + 2\nu_{12})(-1 + 2\nu_{12})}{4 (10 - 3\mu_{\rm eff})^2 \nu_1 (1 + \nu_1)(-1 + 2\nu_1) \nu_2 (1 + \nu_2)(-1 + 2\nu_2)} \times \\[5pt]
& \left[ 58 + 98\nu_1^3\nu_2 + (3 - 91\nu_2)\nu_2 + 7\nu_1^2(-13 - 2\nu_2 + 28\nu_2^2) \right. \\
& \quad \left. + \, \nu_1(3 + 2\nu_2(-73 + 7\nu_2(-1 + 7\nu_2))) \right]
\end{aligned}
\end{equation}
As expected, in the standard gravity limit where $\mu_{\rm eff} = 1$, $M_{22}$ reduces to the conventional matrix of $\Lambda$CDM \cite{Chudaykin:2020aoj}.

A similar procedure can be applied to the $P_{13}(k)$ term, whose expansion can be written as follows:
\begin{equation}
\bar{P}_{13}(k) = 6 P_{\rm L}(k) \sum_{m_1} c_{m_1} \sum_{n_1, n_2} f_{13}(n_1, n_2) \, k^{-2(n_1 + n_2)} \int_{\mathbf{q}} \frac{1}{q^{2\nu_1 - 2n_1} |\mathbf{k} - \mathbf{q}|^{-2n_2}} \, .
\end{equation}
Solving the momentum integral, this expression can be further simplified to:
\begin{equation}
\bar{P}_{13}(k) = k^3 P_{\rm L}(k) \sum_{m_1} c_{m_1} k^{-2\nu_1} \cdot M_{13}(\nu_1) \, ,
\end{equation}
where the vector $M_{13}(\nu_1)$ is given by
\begin{equation}
M_{13}(\nu_1) = \frac{(1 + 9\nu_1) 3\mu_{\rm eff}}{4} \frac{\tan(\nu_1 \pi)}{8\pi\nu_1 (70 - 31\mu_{\rm eff}+ 3\mu_{\rm eff}^ 2)  (-6 + 5\nu_1 + 5\nu_1^{2} - 5\nu_1^3 + \nu_1^4)}  .
\end{equation}
Once again, in the standard gravity limit ($\mu_{\rm eff} = 1$), we recover the standard $M_{13}$. With these two expressions for $M_{22}$ and $M_{13}$ at hand, we obtain an efficient scheme to evaluate the integrals of Eqs.~\eqref{P22new} and ~\eqref{eq:P13mond}. That is the one-loop matter power spectrum in MOND-like modifications to gravity, which can be readily implemented in numerical codes.

%At this point we can follow the same procedure done before in \cite{Silva:2025bnn} to evaluate the integral that corrects the $P_{13}$ term in the MOND-like model. We orientate \(\mathbf{k}\) along the \(z\)-axis and introducing \(\mu=\hat{\mathbf{k}}\cdot\hat{\mathbf{q}}\). In that manner it is able to reproduce the following integration:
%\begin{equation}
%\int d^3q\;\mathcal{I}(k,q,\mu) P_L(q)
%= 2\pi\int dq\,q^2 P_L(q)\int_{-1}^{1} d\mu\; \mathcal{I}(k,q,\mu).
%\end{equation}
%where $\mathcal{I}(k,q,\mu)$ is the integrate of equation (\ref{DP13}) after be simetrized and explicited applied the specific wave vectors.

%After doing all that process the final expression for $\Delta P_{13}$ takes the form:
%\begin{equation}    \Delta P_{13}(k) = \frac{2 \pi}{21 - 3\mu_{\rm eff}} \int dq\ q^2\ P_L (q) \ f(k,q)
%\end{equation}
%where the f function is:
%\begin{equation}
%\begin{aligned}
%f(k,q) = & \, \frac{\mu_{\rm eff} -1}{7(10 - 3\mu_{\rm eff})} \left( 1208 - 96\frac{k^2}{q^2} - 32\frac{(k+q)^2}{q^2} - \frac{420k^2 + 980q^2}{(k+q)^2} - 40\frac{k^4}{q^2(k+q)^2} \right. \\[1.5ex]
%& \left. - \frac{420k^2 + 980q^2}{(k-q)^2} - 56\frac{(k+q)^2}{(k-q)^2} - 40\frac{k^4}{q^2(k-q)^2} - 16\frac{k^2 (k+q)^2}{q^2(k-q)^2} \right)
%\end{aligned}
%\end{equation}
%this extra function encodes the deaperture caused by a MOND-like gravity to the standard $\Lambda$CDM scenario. As expected, when $\mu_{\rm eff} =1$ the extra function vanishes and in this case $\Delta P_{13}=0$. 

In summary, in the mildly nonlinear regime and in real space, the full one-loop matter power spectrum, with the EFT framework, in the presence of a MOND-like modification to gravity can be expressed as:

%(As we probe the mildly non-linear regime, incorporating corrections from the EFTofLSS framework is essential. This regularization is achieved through Wilsons coeficients, namely counterterms, which naturally absorb the UV ($q >> k$) cutoff dependence, thereby canceling the small-scale divergence of the $P_{\rm \space 1-loop}$ contribution. Notably, the summation of the leading IR ($k >> q$) -divergent contributions yields a complete cancellation, ensuring an IR-finite result \cite{Ivanov:2022mrd}. This cancellation is a direct consequence of the Galilean invariance underlying the equations of motion and, crucially, holds holds true for all subsequent subleading IR divergences as well. The UV limit of $P_{22}$ matches the structure of the stochastic term and can be safely neglected at one-loop order \cite{}. Conversely, the one-loop EFTofLSS counterterm arises from absorbing the cutoff dependence inherent in the UV limit of the $P_{13}$ integral \cite{}. We retain the usual expression of the one-loop counterterm such that the total one-loop power spectrum within the EFTofLSS framework for the MOND-like modification to gravity is given by:
%(One of the primary advantages of adopting the methodology presented in [] is that the asymptotic behavior of the one-loop integrals retains its standard form. In this manner, the counterterm inherits the exact same functional expression as in standard $\Lambda\text{CDM}$. Consequently, the total one-loop power spectrum within the EFTofLSS framework for the MOND-like modification to gravity is given by)%

\begin{equation} \label{eft}
    P^{\rm Mond}_{\rm 1-loop, \ EFT}(k) = P_L(k) + \left(\frac{7}{10 - 3\mu_{\rm eff}} \right)^2 P_{22}^{\rm std}(k) + \frac{18}{21-3\mu_{\rm eff}}P_{13}^{\rm std}(k) + \Delta P_{13}(k) -2c_s k^2P_L(k).
\end{equation}
where $c_s$ denotes the effective speed of sound, which accounts for unresolved small-scale physics. It is worth emphasizing that higher-order loop corrections require a larger set of counterterms. Therefore, the EFT correction in Eq.~(\ref{eft}) accounts specifically for the regularization of the one-loop contribution.

\begin{figure}[htpb!]
    \centering
    \includegraphics[scale=0.475]{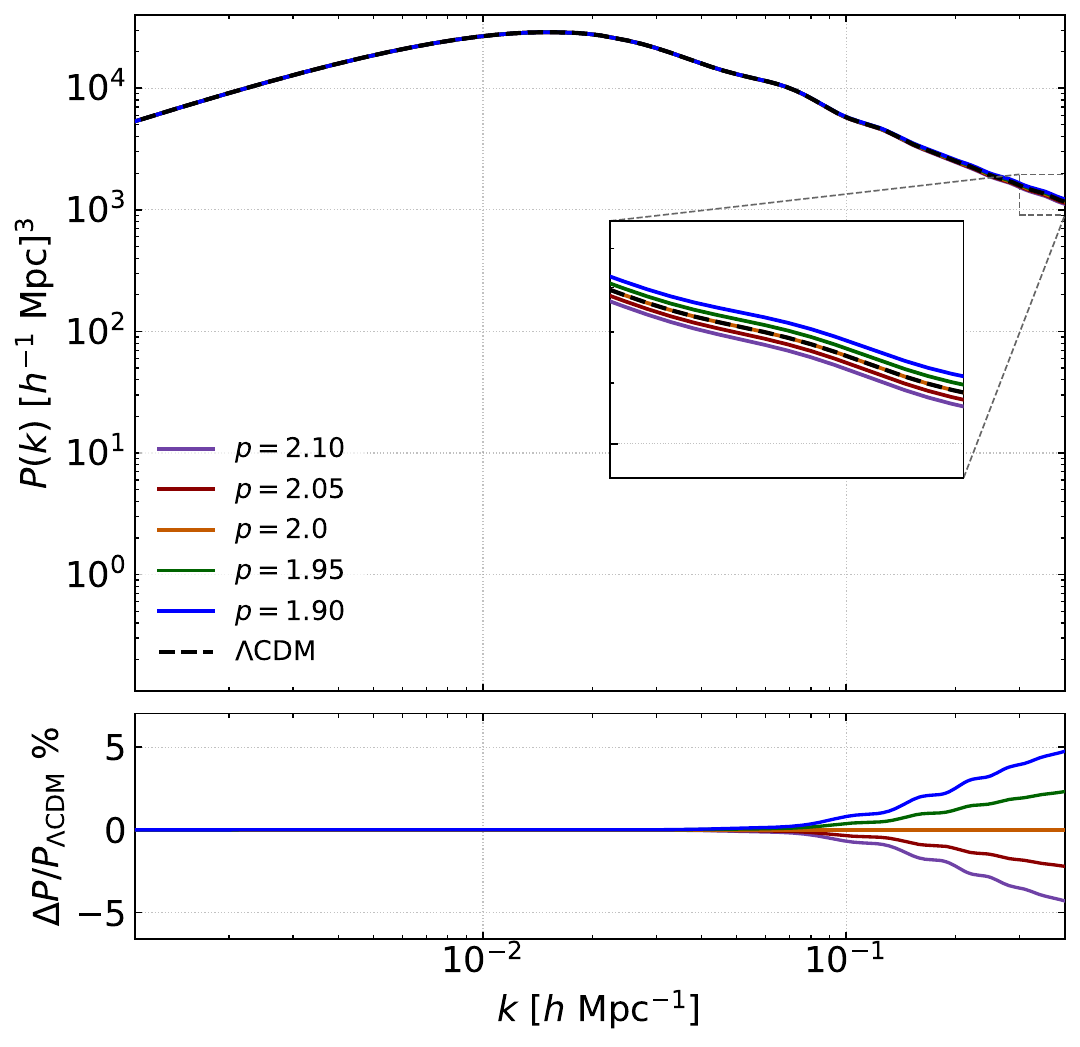}
    \includegraphics[scale=0.475]{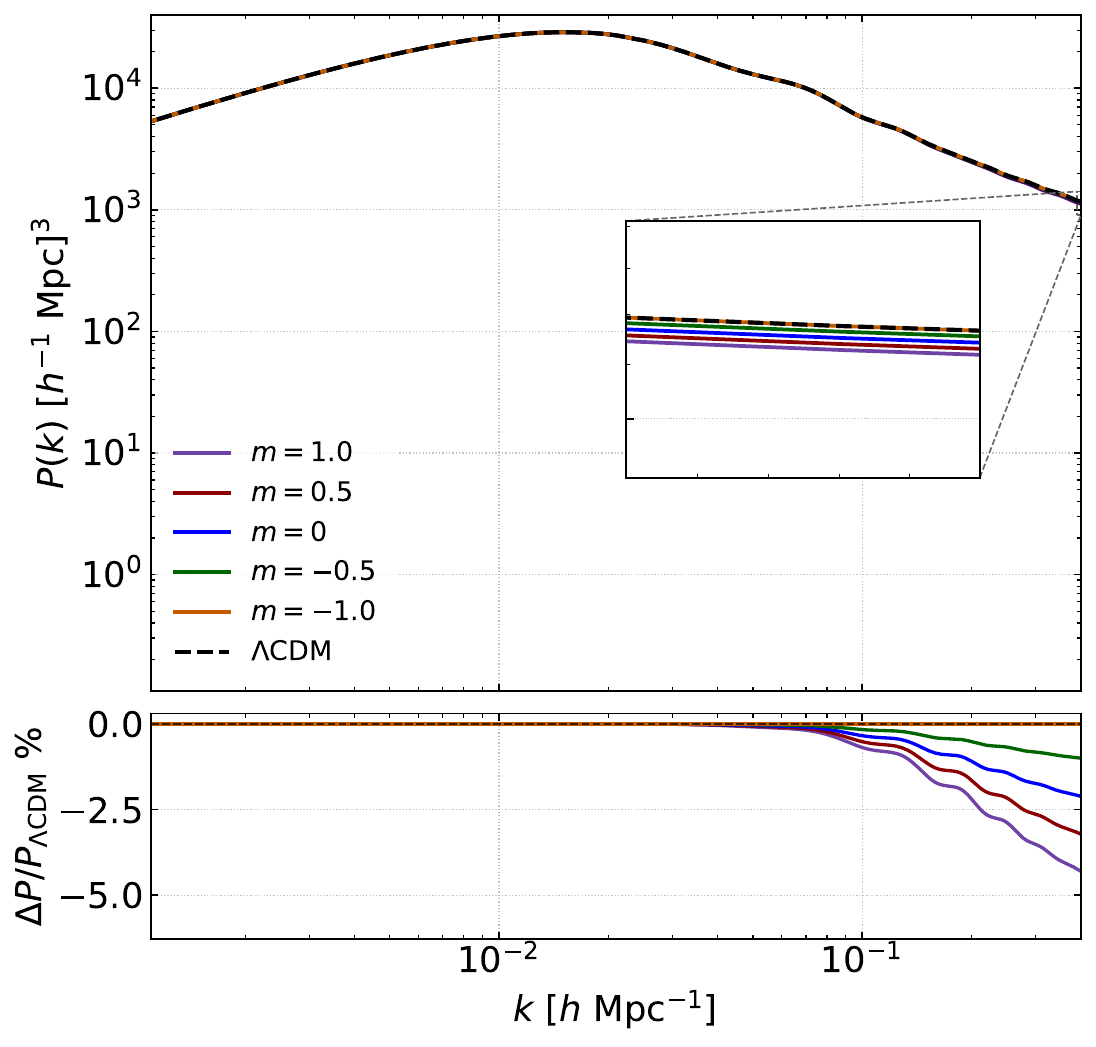}
    \caption{\textit{Left panel:} One-loop matter power spectrum obtained for different values of $p$ (see legend). The black dashed curve corresponds to the best-fit $\Lambda$CDM model. The lower subpanel shows the relative difference of each model with respect to the $\Lambda$CDM prediction. \textit{Right panel:} Same as the left panel, but with $p=2.10$ fixed while varying only $m$.}
    \label{fig:p(k)}
\end{figure}

For a first qualitative assessment, we plot the matter power spectrum including the perturbative corrections derived above. This is shown in Fig.~\ref{fig:p(k)}, which highlights the impact of the MOND-like one-loop contributions on the scale dependence of the matter power spectrum. Since we are analyzing MOND-inspired gravitational deformations implemented on top of a $\Lambda$CDM background, the matter power spectrum remains unchanged on large scales. As we move through the mildly non-linear regime ($k \approx 10^{-1} \, h\,\text{Mpc}^{-1}$), the deviations start to appear, becoming more pronounced toward smaller scales. 

We find that larger values of $p$ lead to a stronger suppression of the matter power spectrum on nonlinear scales, reaching deviations of up to $5\%$ within the range explored. Physically, this behavior is driven by the effective gravitational coupling $\mu_{\rm eff}$. In general, increasing $p$ decreases $\mu_{\rm eff}$, corresponding to a weaker effective gravitational interaction and, consequently, less efficient structure formation. Conversely, smaller values of $p$ enhance the power spectrum relative to $\Lambda$CDM.

A similar trend is observed when varying $m$, since $m$ is also inversely related to $\mu_{\rm eff}$. However, in the plot, $m$ appears only to suppress $P(k)$. This should not be interpreted as an intrinsic physical effect of $m$, which can also lead to amplitudes larger than those predicted by $\Lambda$CDM. Rather, this behavior results from fixing $p=2.10$, which already leads to a suppressed spectrum. Since $m$ can effectively act as a tuning parameter for the effect of $p$, its variation follows the same trend in this particular case. Consistently, the $\Lambda$CDM limit is recovered for $p=2$. For $m=0$, the $\Lambda$CDM fit is not recovered because we fix $p=2.10$ to probe deviations from the $\Lambda$CDM prediction.

When extending the EFTofLSS framework to full-shape analyses of galaxy surveys, several observational and nonlinear effects must be systematically included, such as IR resummation, galaxy bias, and redshift-space distortions. The present study represents an initial methodological step toward incorporating physics beyond $\Lambda$CDM, particularly MOND-like models, into analyses of mildly nonlinear clustering. A complete treatment of these effects, along with full-shape parameter constraints, is deferred to future work due to the significant theoretical and computational developments required.

%\appendix
\subsection{Time dependence}
\label{sec:appA}

The EdS approximation is commonly employed in full-shape analyses of modified gravity scenarios \cite{Ishak:2024jhs, Chudaykin:2024gol}. However, this approximation can break down when the background dynamics deviates from $\Lambda$CDM; see, for example, Refs.~\cite{Piga:2022mge, DESI:2026haa, Noriega:2022nhf, Aviles:2021que}. In this appendix, we carefully derive the differential equations governing the perturbation kernels beyond the EdS approximation. In our case, however, an additional time dependence arises from the effective gravitational constant, which further modifies the evolution of the kernels.

Following the standard approach, explicitly shown in \cite{Zheng:2025owb}, we express the cosmological fluid equations in terms of the rescaled velocity divergence $\Theta \equiv - \ \nabla \cdot \mathbf{v}/\mathcal{H}$ and adopt the number of e-folds $N = \ln a$ as the time variable. Denoting derivatives with respect to $N$ by a prime ($\partial_N \equiv {}'$), the continuity and Euler equations in Fourier space can be written as:

\begin{equation}
 \delta'(\mathbf{k}, N) - \Theta(\mathbf{k}, N) = 
 \int _{q_1,q_2} \, \delta_D(\mathbf{k} - \mathbf{q}_1 - \mathbf{q}_2) \, \alpha(\mathbf{q}_1, \mathbf{q}_2) \, \Theta(\mathbf{q}_1, N) \, \delta(\mathbf{q}_2, N),
\end{equation}
\begin{equation}
    \Theta'(\mathbf{k}, N)
+R(N) \ \Theta(\mathbf{k}, N) 
- \frac{3}{2} \Omega_m(N)  \mu_{\rm eff}(\mathbf{k},\mu,N) \delta(\mathbf{k}, N) 
= \nonumber \\
 \int _{q_1,q_2} \, \delta_D(\mathbf{k} - \mathbf{q}_1 - \mathbf{q}_2) 
\, \beta(\mathbf{q}_1, \mathbf{q}_2) \Theta(\mathbf{q}_1, N)\Theta(\mathbf{q}_2, N),
\end{equation}
where $R$ is defined as $R(N) \equiv 2 + H'/H$.
The perturbative approach of SPT allows us to expand the non-linear density and velocity divergence fields order-by-order in terms of the linear density field $\delta_1$ $\delta(\mathbf{k}, N) = \sum_{n=1}^{\infty} \delta_n(\mathbf{k}, N),$ and $\Theta(\mathbf{k}, N) =  \sum_{n=1}^{\infty}\Theta_n(\mathbf{k}, N)$
such that at the $n$-th order, the solutions can be written as a convolution of $n$ linear fields, weighted by the kernels $F_n$ and $G_n$:
\begin{equation}
\label{eq:appA}
\delta_n(\mathbf{k}, N) = \int _{q_1 \cdots q_n} \delta_D\left(\mathbf{k} - \sum_{i=1}^n \mathbf{q}_i\right)\, F_n(\mathbf{q}_1, \ldots, \mathbf{q}_n, N) \, \ \delta_1(\mathbf{q}_1, N) \ \cdots \ \delta_1(\mathbf{q}_n, N),
\end{equation}
and
\begin{equation}
\label{eq:appB}
\Theta_n(\mathbf{k}, N) = \int _{q_1 \cdots q_n} \delta_D\left(\mathbf{k} - \sum_{i=1}^n \mathbf{q}_i\right)\, G_n(\mathbf{q}_1, \ldots, \mathbf{q}_n, N) \, \ \delta_1(\mathbf{q}_1, N) \ \cdots \ \delta_1(\mathbf{q}_n, N),
\end{equation}
at linear order ($n=1$), it is straightforward to check that the kernels trivially reduce to $F_1 = 1$ and $G_1 = f(k)$. In second order the Euler and continuity equations after symmetrization take the form:
\begin{equation}
 \delta_2'(\mathbf{k}, N) -  \Theta_2(\mathbf{k}, N) = 
\frac{1}{2}\int _{q_1,q_2} \, \delta_D(\mathbf{k} - \mathbf{q}_1 - \mathbf{q}_2) \Big[\alpha(\mathbf{q}_1, \mathbf{q}_2) f(q_1) + \alpha(\mathbf{q}_2, \mathbf{q}_1)f(q_2) \Big] \, \delta_1(\mathbf{q}_1, N) \, \delta_1(\mathbf{q}_2, N),
\end{equation}
\begin{equation}
\Theta_2'(\mathbf{k}, N)
+ R \Theta_2(\mathbf{k}, N) 
- \frac{3}{2} \Omega_m  \mu_{\rm eff}(\mathbf{k},\mu,N)\delta_2(\mathbf{k}, N) 
= \nonumber \\
 \int _{q_1,q_2} \, \delta_D(\mathbf{k} - \mathbf{q}_1 - \mathbf{q}_2) 
\, \beta(\mathbf{q}_1, \mathbf{q}_2) \delta_1(\mathbf{q}_1, N)\delta_1(\mathbf{q}_2, N),
\end{equation}
additionally, we introduce the definition
\begin{equation}
    S(k, N) \equiv \frac{3}{2} \, \Omega_m(N) \, \mu_{\rm eff}(k,N) \, .
\end{equation}
The main distinguishing feature of our analysis lies in the explicit time dependence of $S(k,N)$, sourced by the effective coupling $\mu_{\rm eff}(k,N)$, as defined in eq.~\eqref{coupling_mond}. 

Substituting the results of Eqs.~\eqref{eq:appA} and \eqref{eq:appB} into the second-order Euler and continuity equations yields a coupled system of equations for the second-order kernels, $F_2$ and $G_2$:
\begin{gather}
    F_2' + F_2\left(f(q_1) + f(q_2)\right) - G_2 = \frac{1}{2}\left[\alpha(\mathbf{q_1},\mathbf{q_2})f(q_1) + \alpha(\mathbf{q_2},\mathbf{q_1})f(q_2)\right] \, , \label{eq:F2_system} \\[8pt]
    G_2' + G_2\left(f(q_1) + f(q_2)\right) + RG_2 - S(k)F_2 = \beta(\mathbf{q_1},\mathbf{q_2})f(q_1) f(q_2) \, . \label{eq:G2_system}
\end{gather}

Combining Eqs.~\eqref{eq:F2_system} and \eqref{eq:G2_system}, and using $f'(q_i) = S(q_i) - Rf(q_i) - f(q_i)^2$ , we obtain a second-order differential equation for the second order density field kernel $F_2$:
\begin{align}
    F_2'' &+ 2\left(f(q_1) + f(q_2) + \frac{R}{2}\right) F_2' + \Big[2f(q_1) f(q_2) + S(q_1) + S(q_2) - S(k)\Big] F_2 \nonumber \\
    &= \frac{1}{2}\Big[\alpha(\mathbf{q_1},\mathbf{q_2}) S(q_1) + \alpha(\mathbf{q_2},\mathbf{q_1}) S(q_2)\Big] + \frac{1}{2}f(q_1) f(q_2) \Big [\alpha(\mathbf{q_1},\mathbf{q_2}) + \alpha(\mathbf{q_2},\mathbf{q_1})\Big ] + \beta(\mathbf{q_1},\mathbf{q_2}) f(q_1) f(q_2) \, . \label{eq:F2_diff}
\end{align}

Once Eq.~\eqref{eq:F2_diff} is solved for $F_2$, the second-order velocity divergence kernel $G_2$ follows directly by inverting the continuity equation, Eq.~\eqref{eq:F2_system}:
\begin{equation}
    G_2 = F_2' + F_2\left(f(q_1) + f(q_2)\right) - \frac{1}{2}\left[\alpha(\mathbf{q_1},\mathbf{q_2})f(q_1) + \alpha(\mathbf{q_2},\mathbf{q_1})f(q_2)\right] \, . \label{eq:G2_solution}
\end{equation}

This closes the system of second-order kernels: $F_2$ is obtained by numerically integrating Eq.~\eqref{eq:F2_diff} with the appropriate initial conditions, while $G_2$ is subsequently reconstructed from Eq.~\eqref{eq:G2_solution}. In this way, the additional time dependence sourced by the effective coupling $\mu_{\rm eff}(k,N)$ is consistently propagated to both kernels, distinguishing our approach from the standard treatment. It is particularly noteworthy that, in the standard $\Lambda$CDM EdS scenario, all time derivatives vanish and the relevant functions reduce to $S(k) = 3/2$ and $f(k) = 1$. With this choice, Eq.~\eqref{eq:F2_diff} simply reduces to $F_2^{s} = \frac{5}{7}\,\alpha^{s} + \frac{5}{7}\,\beta$, recovering the standard EdS kernels.

A full numerical implementation of Eqs.~\eqref{eq:F2_diff} and \eqref{eq:G2_solution}, consistently retaining this time dependence throughout the analysis, remains numerically challenging, particularly when propagating it to the redshift-space power spectrum multipoles, and is therefore beyond the scope of the present work. Nevertheless, the derivation presented here provides the necessary theoretical framework for such an implementation. Without loss of generality, this derivation should be regarded as a general demonstration of the formalism rather than a useful prescription for numerical applications to simulations or real-data analyses.

%, and we expect this extension to play an important role in tightening constraints on MOND-like models with upcoming full-shape galaxy clustering data.

\section{Conclusion}
\label{sec:end}

In this work, we test a generalized MOND-like framework in which modifications to gravity are described by an effective gravitational coupling, $\mu_{\rm eff}$. The model approaches general relativity at high accelerations but allows deviations at low accelerations regimes. We study its effects on cosmic structure growth, from the linear regime to mildly nonlinear scales, and compare its predictions with current large-scale structure observations.

To detail our methodology, we investigate the observational viability of a MOND-like modified-gravity framework in which an effective gravitational coupling alters the dynamics of cosmic structure formation. We first derive the corresponding linear structure formation equations within the generalized Poisson equation framework. We then extend the analysis into the mildly nonlinear regime through the one-loop calculations, deriving the corresponding corrections to the matter power spectrum. Our main results are:

\begin{itemize}
  \item We observe that in this new framework, the effective mass $\mu_{\text{eff}}$ generates significant variations in growth quantities (such as $f(z)$ and $f \sigma_8(z)$), which can be explored in future work  with CMB data to assess the level of tension at $S_8$ and how it affects the concordance between late- and early-time probes.

  \item For $m=1$, the data show a mild preference for $p>2$, with $p\simeq2.1$ across the data combinations considered. This corresponds to a suppressed effective gravitational coupling, $\mu_{\rm eff}<1$, and a lower growth rate than predicted by $\Lambda$CDM, particularly at low redshift.

  \item Allowing $m$ to vary freely retains the preference for $p>2$, with $p\simeq2.16$ for the full Growth+DESI-DR2+U3 combination. The parameter $m$ remains poorly constrained because variations in its redshift dependence can be partially compensated by changes in $p$, $\Omega_m$, and $S_8$. 
  %Current LSS data therefore have limited sensitivity to the time evolution of the gravitational coupling.

  \item On nonlinear scales, variations in the effective gravitational coupling can significantly modify $P_{\rm NL}(k)$, providing a new observational signature for probing MOND-like modifications of gravity. We will explore these consequences in future work
  
%  at small scales. %Future high-precision observations could test this regime.
\end{itemize}

Finally, although negative values of $\Delta\chi^2_{\rm min}$ indicate a marginal improvement in the fit, the $\Lambda$CDM limit, $p=2$, remains consistent with the data at the $1\sigma$ level for every case considered. The MOND-like framework is a viable and physically motivated alternative to standard gravity, but it currently has no statistically significant advantage over $\Lambda$CDM. High-precision surveys, particularly at low redshift and on nonlinear scales data information, could help break parameter degeneracies and test this class of modified-gravity models.

This work opens new avenues for investigating whether MOND-like models remain viable on cosmological scales. Several natural extensions follow from the framework developed here. First, a full treatment of the mildly nonlinear regime should be pursued, incorporating a proper galaxy bias prescription and redshift-space distortion effects, in order to confront the model with full-shape clustering data. Second, a relativistic, covariant formulation of the model should be developed, allowing for a consistent modification of the background cosmology and enabling direct comparisons with CMB data. Together, these extensions would provide a more stringent and comprehensive test of MOND-like scenarios across the full range of cosmological scales, complementing the galactic-scale successes that originally motivated these theories, and will be done in future works.

\begin{acknowledgments}
E.S. received support from the CAPES scholarship. R.C.N. thanks the financial support from the Conselho Nacional de Desenvolvimento Científico e Tecnologico (CNPq, National Council for Scientific and Technological Development) under the project No. 304306/2022-3, and the Fundação de Amparo à Pesquisa do Estado do RS (FAPERGS, Research Support Foundation of the State of RS) for partial financial support under the project No. 23/2551-0000848-3.
\end{acknowledgments}

\bibliographystyle{apsrev4-1}
\bibliography{references}

\end{document}